\documentclass[pra,10pt,twocolumn,groupedaddress,superscriptaddress]{revtex4-1} 

\usepackage[dvips]{graphicx}
\usepackage{subfigure}
\usepackage{amsmath}
\usepackage{amssymb}
\usepackage{bm}

\def\Eq{Eq. \hspace{-0.1cm}}
\def\Eqs{Eqs. \hspace{-0.1cm}}
\def\Fig{Fig. \hspace{-0.1cm}}
\def\Figs{Figs. \hspace{-0.1cm}}
\def\Ref{Ref. \hspace{-0.1cm}}
\def\Refs{Refs. \hspace{-0.1cm}}
\def\be{\begin{equation}}
\def\ee{\end{equation}}
\def\bea{\begin{eqnarray}}
\def\eea{\end{eqnarray}}
\def\ds{\displaystyle}
\newcommand{\bra}[1]{\left\langle #1 \, \right|}
\newcommand{\ket}[1]{\left| #1 \right\rangle}
\newcommand{\braket}[2]{\left\langle #1 \, | #2 \right\rangle}

\begin{document}

\title{Numerical simulation of a multi-level atom interferometer}

\author{B. Barrett}
\affiliation{Department of Physics \& Astronomy, York University, Toronto, ON M3J 1P3}

\author{I. Yavin}
\affiliation{Center of Cosmology \& Particle Physics, New York University, New York, NY 10003}

\author{S. Beattie}
\author{A. Kumarakrishnan}
\affiliation{Department of Physics \& Astronomy, York University, Toronto, ON M3J 1P3}

\bibliographystyle{apsrev4-1}


\begin{abstract}
We present a comprehensive numerical simulation of an echo-type atom interferometer. The simulation confirms a new theoretical description of this interferometer that includes effects due to spontaneous emission and magnetic sub-levels. Both the simulation and the theoretical model agree with the results of experiments. These developments provide an improved understanding of several observable effects. The evolution of state populations due to stimulated emission and absorption during the standing wave interaction imparts a time-dependent phase on each atomic momentum state. This manifests itself as an asymmetry in the signal shape that depends on the strength of the interaction as well as spontaneous emission due to a non-zero population in the excited states. The degree of asymmetry is a measure of a non-zero relative phase between interfering momentum states.
\end{abstract}

\maketitle

\section{Introduction}

In recent years, atom interferometers (AIs) have become an invaluable tool for a variety of experiments related to precision measurements and inertial sensing using cold atoms \cite{Chu6, Chu3, Chu1, Kumar2, Phillips, Pritchard2, Clade, Chu4, Kasevich1, Kasevich2, Lamporesi, Chu5, Beattie2}.

The echo-type AI used in this work \cite{Kumar2, Beattie2, Kumar1, Beattie1, Beattie3, Prentiss2} functions on the basis of phase modulation of the atomic wave function due to the interaction with standing wave (sw) pulses. Although this phase modulation is directly connected to the recoil energy, its functional form has a complicated dependence on a number of mechanisms---such as the dynamic population of magnetic sub-levels, phase shifts due to spontaneous and stimulated processes and the excitation of multiple momentum states. Some of these mechanisms have been studied in previous work \cite{Kumar1, Beattie3, Gosset, Beattie1}, but many aspects of the AI are not fully understood. In this work, we address all of these effects on the basis of a comprehensive numerical simulation and an improved analytical model.

The simulation presented here successfully models data using measured experimental parameters as inputs and completes our understanding of a broad class of observable effects. One such effect is an asymmetry in the shape of signals produced by the AI, which is a manifestation of a non-zero relative phase difference between interfering momentum states. We show that the level of asymmetry is related to the amount of spontaneous emission occurring during sw excitations, as well as stimulated processes such as Rabi flopping.

The analytical model of the AI accounts for effects due to both magnetic sub-levels and spontaneous emission. By including these effects, we avoid the need for a phenomenological model that was previously used in measurements of the atomic recoil frequency \cite{Beattie3}. Using this theoretical treatment, we are also able to estimate magnetic sub-level populations in the experiment.

Although the analytical model provides an improved understanding of the response of the AI, the model is still limited to short pulses such that the motion of the atom during the interaction can be ignored (Raman-Nath regime). Additionally, the model is limited to detunings ($\Delta$) large compared to the Rabi frequency ($|\Delta| \gg \Omega_0$) and the spontaneous emission rate ($|\Delta| \gg \Gamma$). The analytical calculation also assumes that the excited state adiabatically follows the ground state, eliminating any dynamic exchange of amplitude or phase between states. The decay of the excited state due to spontaneous emission is also accomplished in an approximate manner, since the ground state amplitude is not re-populated by the excited state. As a result of these limitations, the theory fails to model experimental data accurately in several regimes of interest.

Since the simulation numerically solves the equations governing the system, none of the aforementioned limitations apply. The simulation accounts for the motion of atoms and the dynamic evolution of magnetic sub-levels populations during the interaction with sw pulses, which is not a feature of the analytical calculation. In general, this approach allows for a much broader class of phenomena to be studied. Additionally, a precision measurement of the recoil frequency using this technique would involve a detailed study of systematic effects. This justifies the need for an accurate and robust model for the signal.

The rest of the paper is organized as follows. Section \ref{sec:Exp} gives a brief description of the experiment. Section \ref{sec:Theory} reviews the main results pertaining to a theoretical calculation of the signal in two temperature regimes---one reminiscent of Bose-Einstein condensate (BEC) conditions involving one sw pulse, and the other similar to conditions in a magneto-optical trap (MOT), involving two sw pulses. A detailed description of the calculation of the one-pulse recoil signal is given in the Appendix. In \S\ref{sec:Theory}, we also derive an expression for the signal that accounts for magnetic sub-levels, which is used to model experimental data. In \S\ref{sec:Sims}, we describe the theoretical background for the simulations. A discussion of the main results of this work follows in \S\ref{sec:Results}. We present our conclusions in \S\ref{sec:Conclusion} and discuss future directions and applications of this work.

\section{Description of experiment}
\label{sec:Exp}

\begin{figure}[!t]
  \includegraphics[width=0.25\textwidth]{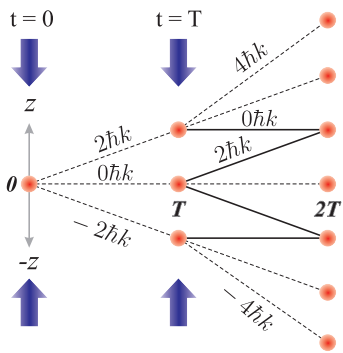}
  \caption{(Color online) Recoil diagram for a time-domain atom interferometer. Center of mass momentum states are shown as red dots. The $1^{\rm{st}}$ sw pulse, applied at $t = 0$, diffracts the momentum states of the atom into multiples of the two-photon recoil momentum $2\hbar k$. The $2^{\rm{nd}}$ sw pulse, applied at $t = T$, splits the momentum states further. Only the $0^{\rm{th}}$ order and the $\pm 1^{\rm{st}}$ order diffractions from each sw pulse are drawn for simplicity. In the vicinity of the echo time, $t = 2T$, there is interference between all orders of momentum states. Our detection scheme is only sensitive to interferences between states that differ by $2\hbar k$. Two pairs of interfering momentum states are shown as solid black lines.}
  \label{fig:RecoilDiagram}
\end{figure}

The AI is used to measure the two-photon atomic recoil frequency $\omega_q = \hbar q^2/2 M$. Here, $M$ is the mass of the atom and $\hbar q = 2 \hbar k$ is the momentum transferred to the atom from counter-propagating laser fields with wavelength $\lambda$ and wavenumber $k = 2\pi/\lambda$. A standing wave laser pulse with off-resonant traveling wave components interacts with a sample of laser cooled $^{85}$Rb atoms (temperature $\mathcal{T} \sim 100$ $\mu$K) at times $t = 0$ and $t = T$. During each pulse, atoms complete several two-photon transitions corresponding to the absorption of a photon from one traveling wave component and stimulated emission into the oppositely directed traveling wave. This results in the diffraction of atoms into a superposition of momentum states separated by $\hbar q$, as shown in \Fig \ref{fig:RecoilDiagram}.

\begin{figure}[!t]
  \includegraphics[width=0.49\textwidth]{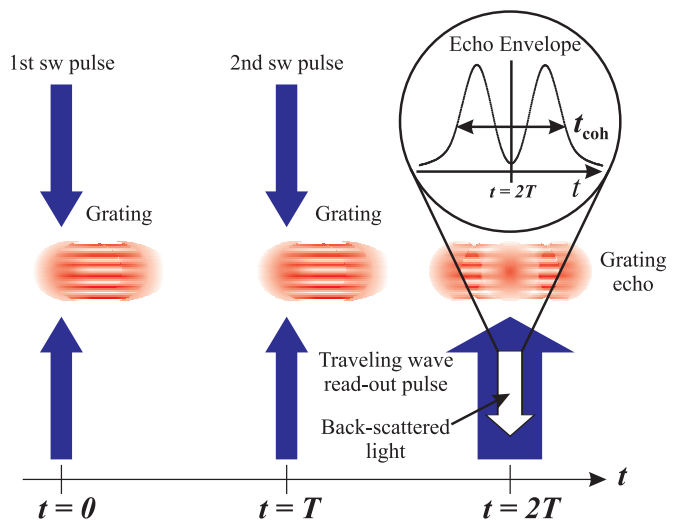}
  \caption{(Color online) Pulse timing diagram for the experiment. After each sw pulse, a modulation in atom density forms and then decays in a time $t_{\rm{coh}} \sim 2/q\sigma_v$ due to Doppler dephasing. At time $t = 2T$, the grating echo forms as a result of the interference between different momentum states. A traveling wave read-out pulse coherently back-scatters light from the grating at this time. The intensity of this light is detected by a photo-multiplier tube.}
  \label{fig:GratingEcho}
\end{figure}

In the absence of spontaneous processes, and assuming both traveling wave components of the standing wave have the same polarization, the interaction returns the atoms to the same ground state magnetic sub-level. The atomic wave function develops a phase modulation on a time scale $\tau_q = \pi/\omega_q \sim 32$ $\mu$s, where $\tau_q$ is referred to as the recoil period. This phase modulation evolves into a spatial and temporal modulation in the atomic density. Due to the finite velocity distribution of the atomic sample, the density grating dephases on a time scale defined by the coherence time $t_{\rm{coh}} \sim 2/q \sigma_v \sim 1$ $\mu$s. Here, $\sigma_v \sim 10$ cm/s is the width of the velocity distribution. Since $t_{\rm{coh}} \ll \tau_q$ for the conditions of our experiment, an echo technique is used to cancel the effect of Doppler dephasing and measure the temporal modulation induced on the density distribution due to atomic recoil. A second sw pulse, applied at $t = T$, causes interference of momentum states in the vicinity of the echo time, $t = 2T$, resulting in a rephasing of the density grating as shown in \Fig \ref{fig:GratingEcho}. The grating contrast is measured by applying a traveling wave read-out pulse (with the same wavelength as the sw pulses) and detecting the intensity of the coherently back-scattered light from the atomic cloud. The density grating has various spatial harmonics in integer multiples of $q$ due to the different orders of interfering momentum states. However, due to the nature of Bragg diffraction, this detection technique is only sensitive to density modulation that has a spatial periodicity of $\lambda/2$ (spatial harmonic $q$)---the smallest grating spacing that can Bragg scatter light of wavelength $\lambda$. Only interfering momentum states that differ by $\hbar q$ can produce such a modulation. The integrated intensity of back-scattered light (referred to as the echo intensity) is proportional to the contrast of the density grating. The echo intensity is a periodic function of the pulse separation, $T$, with period $\tau_q$.

The experiment utilizes $^{85}$Rb atoms cooled in a magneto-optical trap contained within a glass vacuum system. The trap loads $\sim 10^9$ atoms in $\sim 300$ ms. Both the MOT laser beams and the magnetic field gradient are switched off prior to the AI experiment. The MOT B-field gradient is turned off in $\sim 100$ $\mu$s, after which the atoms are cooled in an optical molasses for several milliseconds. Three pairs of coils, one pair along each direction, are used to cancel residual magnetic fields and magnetic field gradients. These coils remain on continuously. Under these conditions, the magnetic field at the time of the echo experiment is canceled at the level of 1 mG over the volume of the trap. The AI excitation pulses are derived from two off-resonant, circularly polarized, traveling wave beams. They are overlapped at the location of the trap to form a standing wave along the vertical direction. The AI beams have a Gaussian intensity profile and are collimated to a diameter of $\sim 2$ cm. All excitation pulses are produced by using digital delay generators to trigger acousto-optic switches. The back-scattered light intensity from the grating is detected using a gated photo-multiplier tube. Measurements of $\omega_q$ are accomplished by measuring the contrast of the grating as a function of the pulse separation, $T$, on a suitably long time scale. A more detailed description of the experiment is given in \Ref \cite{Beattie1}.

\section{Theory}
\label{sec:Theory}

The theoretical expression for the recoil signal was originally derived in \Ref \cite{Kumar2}. The influence of magnetic sub-levels was subsequently included for studies of nano-structures in cold atoms \cite{Kumar1}. The treatment in \Ref \cite{Beattie3} addressed the role of spontaneous emission on the recoil signal. In this work, we include both effects to develop a complete understanding of the properties of the recoil signal. Since theoretical derivations of this signal have been studied previously, we present the details in the appendix. Here, we review the main results required to understand the recoil signal in the one-pulse regime (appropriate for BEC conditions) and the two-pulse regime (used for experiments under MOT conditions). These results are used as a foundation for developing numerical techniques and for testing the accuracy of simulations.

\subsection{One-pulse recoil signal}
\label{sec:Theory-OnePulse}

The one-pulse regime assumes the velocity distribution of the sample is infinitely narrow. The calculation is carried out in three stages. In the first stage, the Schr\"{o}dinger equation is solved for the ground state amplitude of the atomic wave function, $a_g$, using the two-level Hamiltonian for a sw laser pulse of duration $\tau_1$. The Hamiltonian assumes that the sw pulse is short (Raman-Nath regime) such that the motion of the atoms along the axis of the standing wave can be neglected during the interaction. This allows us to ignore the kinetic energy term in the Hamiltonian during the sw pulse. It also assumes that the pulse is far off-resonance ($|\Delta| \gg \Omega_0, \gamma$) such that the excited state is not significantly populated. Here, $\Delta \equiv \omega - \omega_0$ is the detuning of the laser frequency, $\omega$, from the atomic resonance frequency, $\omega_0$, $\Omega_0 \equiv \mu_{eg} E_0/\hbar$ is the Rabi frequency, $\mu_{eg}$ is a dipole matrix element, $E_0$ is the electric-field amplitude of each traveling wave component of the sw and $\gamma = \Gamma/2$, where $\Gamma$ is the spontaneous emission rate.

In the second stage of the calculation, after the pulse has turned off, the atom is allowed to evolve in free space for a time $t$, which results in a modification of the phase of the ground state amplitude. In an experiment involving a BEC, a traveling wave read-out pulse with wavelength $\lambda$ can be applied to the atomic sample and the intensity of the back-scattered electric field can be detected as the signal. The amplitude of the back-scattered electric field is proportional to the $\lambda/2$-periodic component of the atomic density modulation (the $2k = q$ Fourier harmonic) produced by the sw interaction.

The final stage of the calculation requires a computation of the probability density of the ground state, $\rho_g(\bm{r},t)$. The recoil signal is obtained by evaluating the square of the $q$-Fourier harmonic of this probability density.

The Hamiltonian in the field-interaction representation \cite{Berman1} for a two-level atom is
\be
  \label{eqn:H-1}
  \tilde{\mathcal{H}} = \hbar \left(
  \begin{array}{cc} -\Delta - i \gamma & \Omega(\bm{r}) \\ \Omega(\bm{r}) & 0 \\ \end{array} \right),
\ee
where $\Omega(\bm{r}) = \Omega_0 \cos(\bm{k} \cdot \bm{r})$ for a sw laser field. The energy is defined to be zero for the ground state and $-\hbar \Delta$ for the excited state. The $-i \hbar \gamma$ term is a phenomenological constant added to account for spontaneous emission during the interaction, which gives rise to amplitude decay of the excited state. This approach is valid in any open two-level system, but only approximately accounts for spontaneous emission in a closed system since the excited state population is not fed into the ground state (normalization is not preserved). In addition, this Hamiltonian does not account for the atomic recoil due to the photon emission. By using the density matrix approach \cite{Berman2} or resorting to Monte-Carlo wave function techniques \cite{Molmer}, a more complete model of spontaneous emission can be realized.

As shown in the appendix, the amplitude of the back-scattered electric field as a function of the time, $t$, after the sw pulse is
\be
  \label{eqn:E1-Theory}
  \tilde{E}_1(t) \propto -u_1 \sin(\omega_q t - \theta) \big[ J_0( \kappa_1 ) + J_2( \kappa_1 ) \big],
\ee
where $u_1$ is the magnitude of the sw pulse area (\Eq \ref{eqn:u1}), $\omega_q$ is the two-photon recoil frequency, $\theta$ is a parameter associated with spontaneous emission given by
\be
  \label{eqn:theta}
  \theta = \tan^{-1}\left( -\frac{\Gamma}{2\Delta} \right),
\ee
$J_{\nu}(x)$ is the $\nu^{\rm{th}}$ order Bessel function of the $1^{\rm{st}}$ kind and $\kappa_1$ is given by \Eq \ref{eqn:kappa1}. From \Eq \ref{eqn:E1-Theory} it is clear that the field amplitude is proportional to $u_1$ in the small pulse area regime, and is shifted by a phase $\theta$ due to spontaneous emission. This has interesting consequences for the signal shape and will be discussed in detail in \S \ref{sec:Results-SE}.

\begin{figure}[!t]
  \subfigure{
    \label{fig:LowPulseArea-DensityPlot}
    \includegraphics[width=0.23\textwidth]{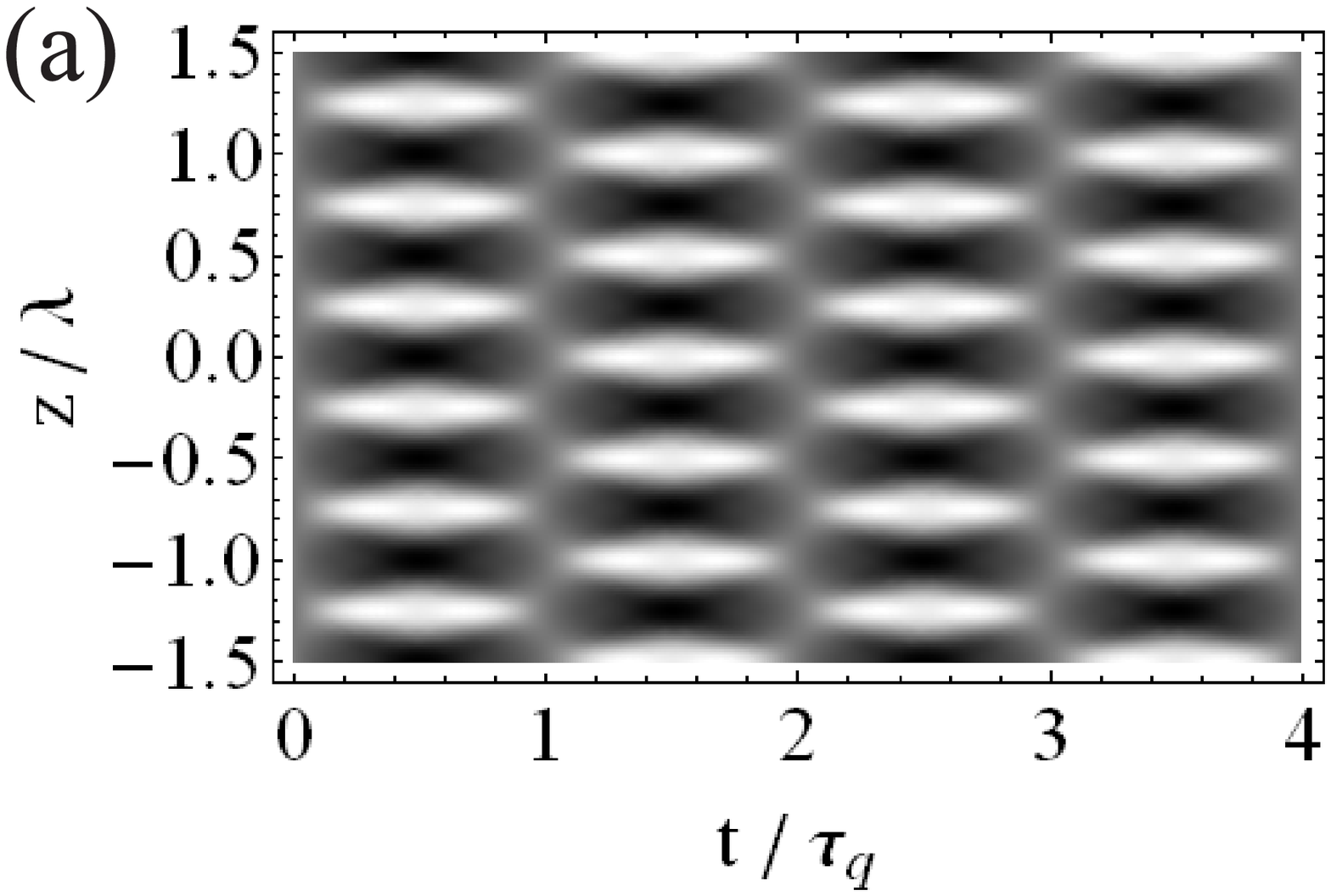}}
  \subfigure{
    \label{fig:HighPulseArea-DensityPlot}
    \includegraphics[width=0.23\textwidth]{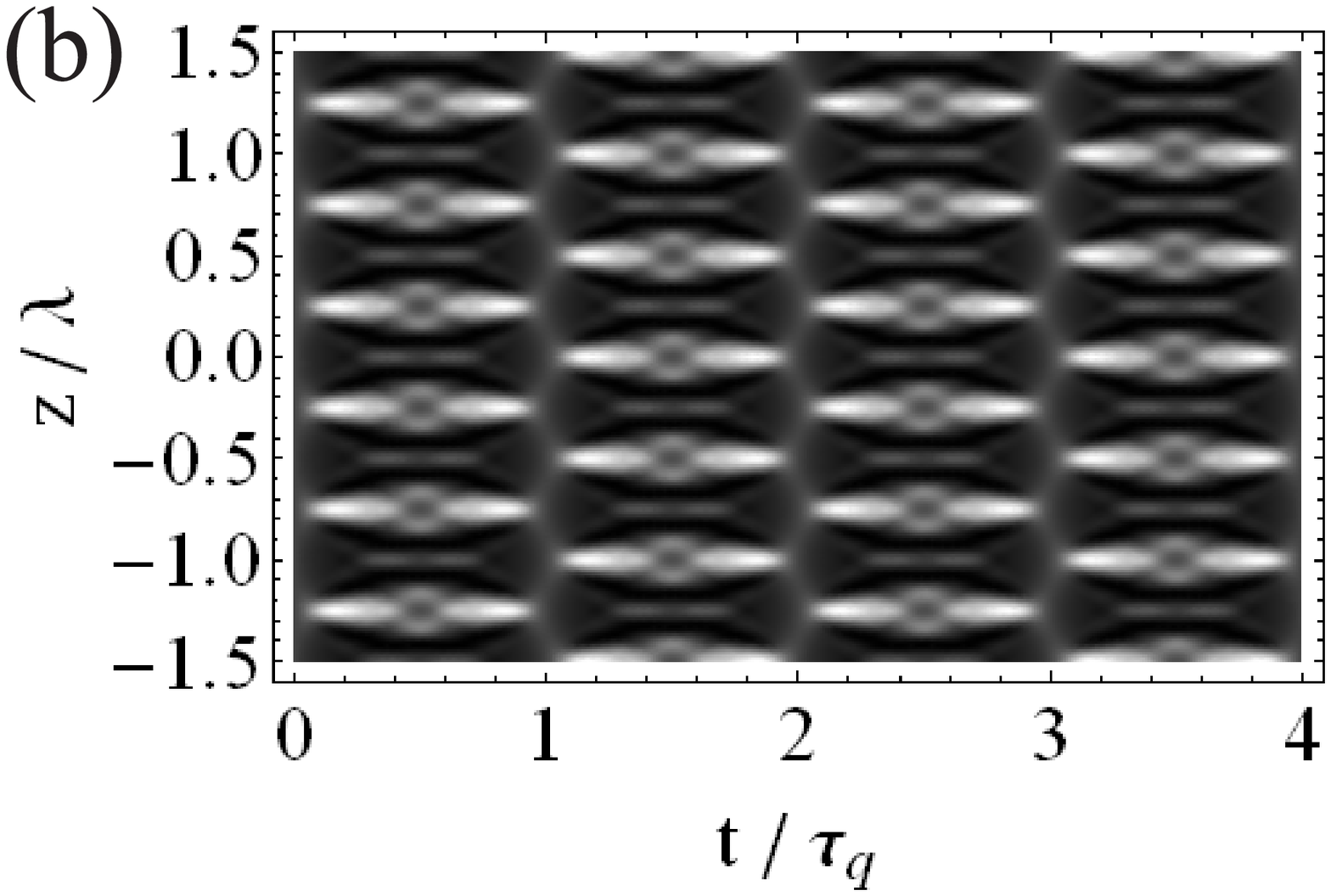}}
  \subfigure{
    \label{fig:LowPulseArea-E1}
    \includegraphics[width=0.23\textwidth]{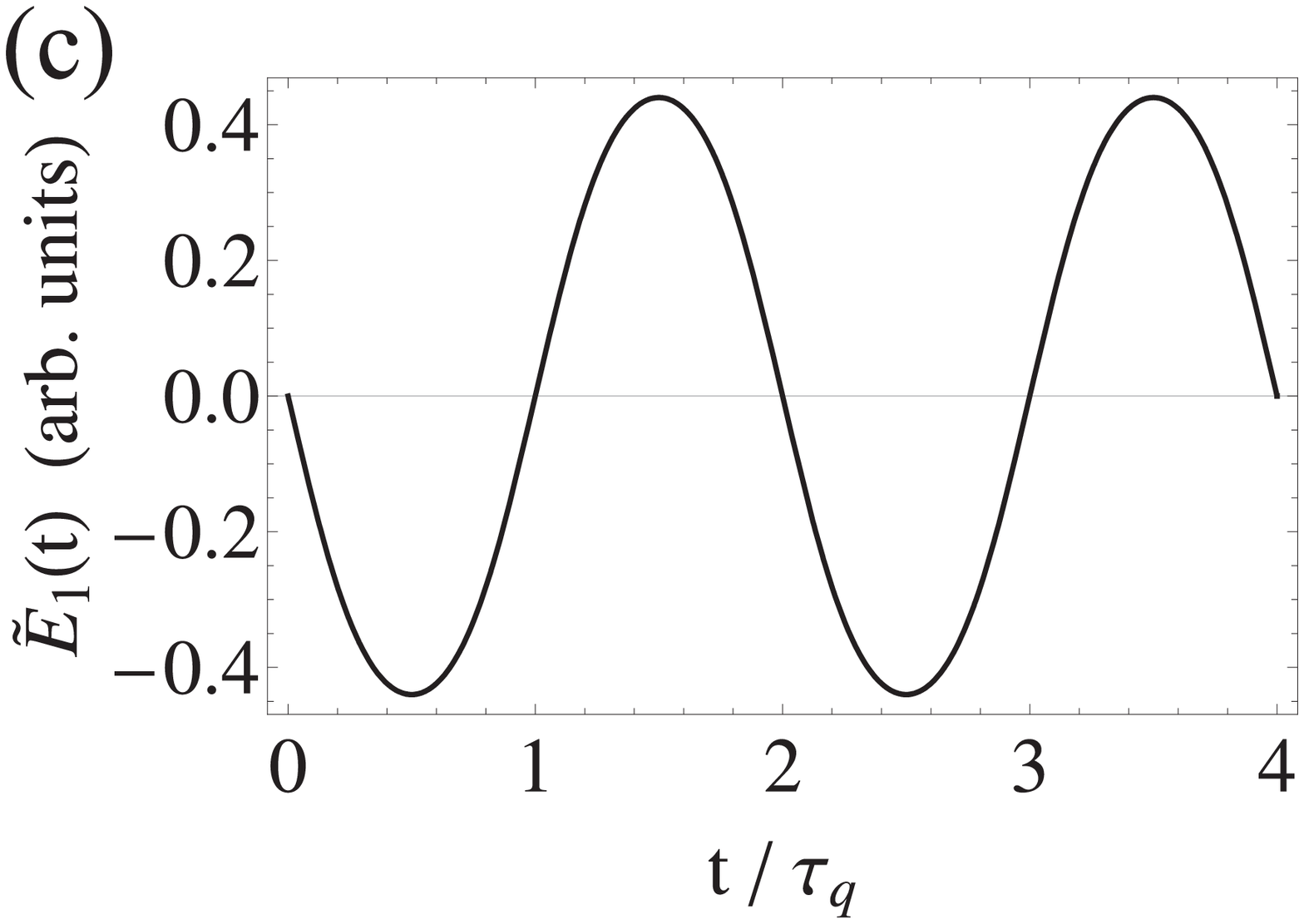}}
  \subfigure{
    \label{fig:HighPulseArea-E1}
    \includegraphics[width=0.23\textwidth]{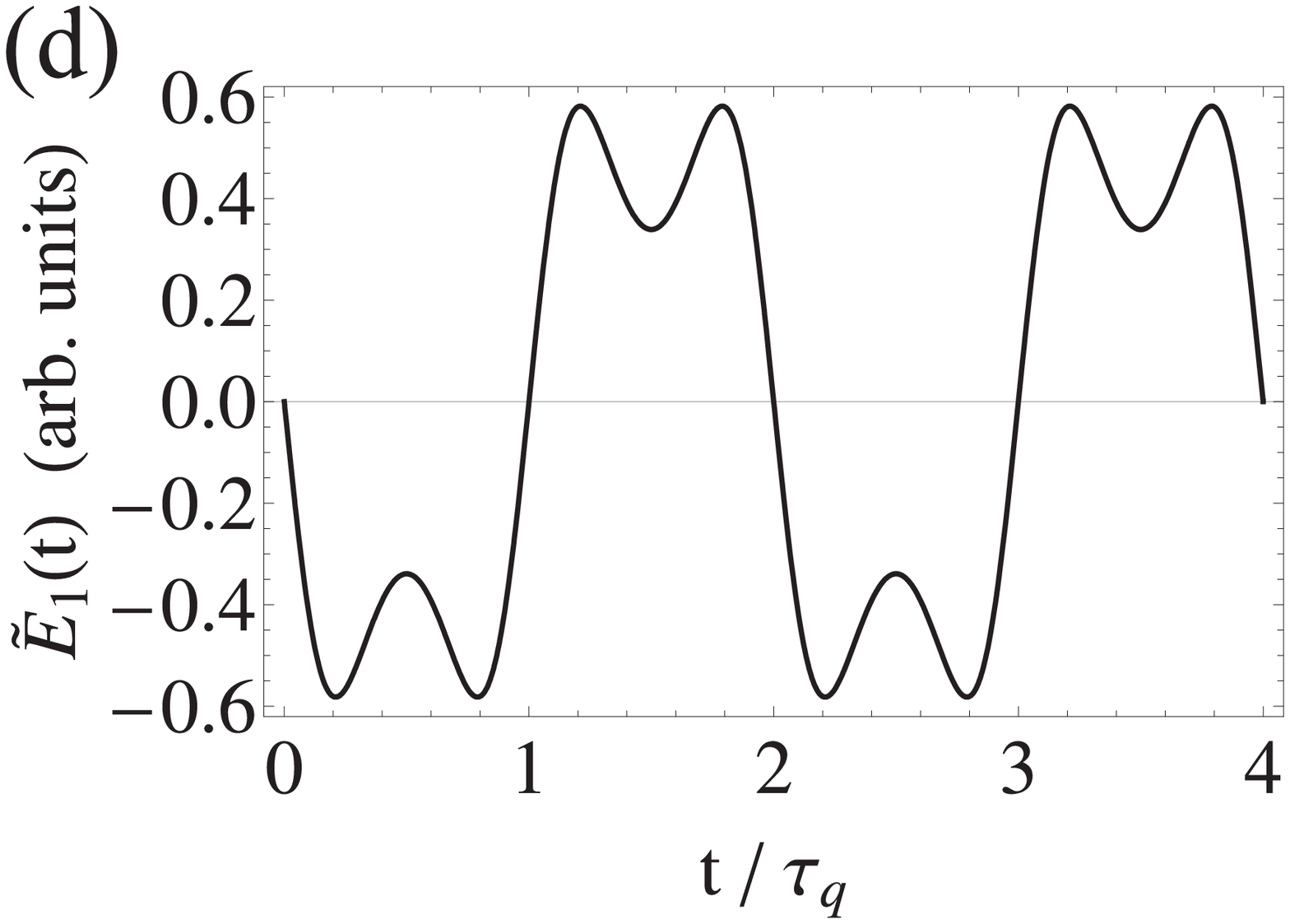}}
  \caption{Density distribution, $\rho_g(z,t)$, and corresponding back-scattered electric field amplitude, $\tilde{E}_1(t)$, for an atomic sample after the interaction with a weak sw pulse (a, c) and a strong sw pulse (b, d). In plots (a) and (b), $z$ is the distance along the sw field. Dark portions correspond to low density while light portions correspond to high density. $\rho_g(z,t)$ shows spatial modulation that is periodic at integer multiples of $\lambda/2$, and shows temporal modulation at integer multiples of twice the recoil frequency, $2\omega_q$, as the pulse area increases. Equation \ref{eqn:rho(r,t)} was used to produce plots (a) and (b) with first pulse area $u_1 = 0.5$ and 1.5 respectively. In plots (c) and (d), \Eq \ref{eqn:E1-Theory} was used with the same respective pulse areas. The zeroes in the contrast of the density modulation correspond to the zeroes in $\tilde{E}_1(t)$. The effects of spontaneous emission were ignored for all plots.}
  \label{fig:OnePulseSignal}
\end{figure}

For sufficiently cold atomic samples ($\sim 10$ nK), $t_{\rm{coh}} \gg \tau_q$ and Doppler dephasing is negligible thereby allowing the temporal modulation in the contrast to be resolved after a single sw pulse \cite{Pritchard2}. The ground state density after the interaction with the sw pulse (\Eq \ref{eqn:rho(r,t)}) is shown in \Figs \ref{fig:LowPulseArea-DensityPlot} and \ref{fig:HighPulseArea-DensityPlot} for low and high pulse areas, respectively, in the absence of spontaneous emission ($\theta = 0$). For small pulse areas, the density is sinusoidally modulated in space with a period $\lambda/2$. The contrast of this spatial modulation oscillates in time with a period $\tau_q$, as shown in \Fig \ref{fig:LowPulseArea-DensityPlot}. This density distribution corresponds to the momentum state $\ket{\bm{p} = 0}$ interfering with the $\ket{\bm{p} = \pm \hbar \bm{q}}$ states. For higher pulse areas, the density distribution becomes more complicated (as shown in \Fig \ref{fig:HighPulseArea-DensityPlot}) due to the presence of additional spatial harmonics from the interference of higher order momentum states. The temporal modulation of the contrast also exhibits higher order harmonics but still primarily oscillates at the fundamental frequency $2\omega_q$.

The amplitude of the back-scattered field, $\tilde{E}_1(t)$ (\Eq \ref{eqn:E1-Theory}), derived from the ground state density $\rho_g(z,t)$ given by \Eq \ref{eqn:rho(r,t)}, is shown in \Figs \ref{fig:LowPulseArea-E1} and \ref{fig:HighPulseArea-E1} for the same pulse areas. The field changes with time due to the oscillations in the contrast of the density distribution. For small $u_1$, only the lowest order momentum states contribute to the signal, and $\tilde{E}_1(t)$ oscillates sinusoidally at frequency $\omega_q$. In general, the states $\ket{n\hbar \bm{q}}$ and $\ket{(n+1)\hbar \bm{q}}$ interfere to produce a $\lambda/2$-spatial modulation with a contrast that oscillates in time with a frequency $2(2n+1)\omega_q$. For higher pulse areas, additional harmonics in the density distribution give rise to the shape of the back-scattered field shown in \Fig \ref{fig:HighPulseArea-E1}. Equation \ref{eqn:E1-Theory} takes into account the interference of any two $p$-states differing by $\hbar \bm{q}$ in momentum, but no higher order interferences---for example, between $\ket{n \hbar \bm{q}}$ and $\ket{(n+2)\hbar \bm{q}}$.

The analytical expression for the one-pulse recoil signal, denoted by $\tilde{s}_1$, is simply the intensity of the back-scattered field: $\tilde{s}_1(t) \propto |\tilde{E}_1(t)|^2$. This quantity is proportional to the contrast of the density modulation.

\subsection{Two-pulse recoil signal}
\label{sec:Theory-TwoPulse}

The theoretical expression of the recoil signal in the two-pulse regime is a simple extension of that in the one-pulse regime. Typically the two-pulse experiment is carried out under MOT conditions where Doppler dephasing becomes important because the velocity distribution of the atoms can no longer be approximated by a delta function.

As shown in the appendix, the standing wave interaction is equivalent to a phase grating in diffractive optics. The calculation for the two-pulse recoil signal involves applying an additional phase grating $e^{i\Theta_2 \cos(\bm{q}\cdot\bm{r})}$, associated with the second sw pulse, to the atomic wave function a time $t = T$ after the first pulse (shown in \Eq \ref{eqn:ag(r,t)}). The atom evolves in free space after the second pulse until a time $t_{\rm{echo}}^{(N)} = (N+1)T$, called the $N^{\rm{th}}$ order echo time, where the Doppler phases from all velocity classes cancel. Here, $N = \eta/\zeta$, where $\eta$ is the momentum difference between interfering $p$-states from the first pulse (in units of $\hbar q$) and $\zeta$ is the equivalent quantity for the second pulse. When averaged over the velocity distribution, the atomic density shows modulation with a non-zero contrast only in the vicinity of these echo times. At all other times, the density is uniform since the modulation has been washed out by Doppler dephasing. In the general case, $N$ is a rational number \cite{Kumar1, Berman1}, but in our case we consider only the first order echo which occurs at time: $t_{\rm{echo}}^{(1)} = 2T$, corresponding to $N = 1$. The density of the atomic sample after the two sw pulses and at the first order echo time is depicted in \Fig \ref{fig:GratingEcho}.

The amplitude of the back-scattered electric field about $t = 2T$ can be shown to be
\begin{align}
\begin{split}
  \label{eqn:E2-Theory}
  \langle \tilde{E}_2(\Delta t, T) \rangle
  & \propto \tilde{E}_1(\Delta t) e^{-(\Delta t/t_{\rm{coh}})^2} \\
  & \times \frac{u_2^2}{2} \sin^2 \big( \omega_q (T + \Delta t) - \theta \big) \\
  & \times \left[ J_0(\kappa_2) + \frac{4}{3} J_2(\kappa_2) + \frac{1}{3} J_4(\kappa_2) \right],
\end{split}
\end{align}
where $T$ is the separation between sw pulses, $\Delta t = t - 2T$ is the time relative to the echo time, $u_2$ is the magnitude of the second pulse area and
\be
  \label{eqn:kappa2}
  \kappa_2 \equiv 2 u_2 \sqrt{\sin(\omega_q (T + \Delta t) - \theta) \sin(\omega_q (T + \Delta t) + \theta)}.
\ee
The $\langle \cdots \rangle$ brackets in \Eq \ref{eqn:E2-Theory} indicate that the back-scattered field has been averaged over the initial velocity distribution, which is assumed to be Maxwellian with a characteristic width $\sigma_v = \sqrt{2 k_B \mathcal{T}/M}$, which is also equal to the most probable speed. The coherence time, $t_{\rm{coh}} = 2/q\sigma_v$, is the time about the echo time for which the back-scattered field is non-zero and is determined solely by $\sigma_v$.

From \Eq \ref{eqn:E2-Theory}, which is equivalent to \Eq 25 in \Ref \cite{Beattie3}, it is clear that for small $u_2$ the basic signal dependence on pulse separation, $T$, is a sinusoidal oscillation at frequency $\omega_q$ that is phase shifted by an amount $\theta$ (\Eq \ref{eqn:theta}). This phase shift is associated with spontaneous emission. As the second pulse area becomes large, the third factor in \Eq \ref{eqn:E2-Theory} has a more significant contribution to the signal. This is associated with the interference of higher order momentum states, as discussed in \S\ref{sec:Theory-OnePulse}.

The two-pulse recoil signal is the integrated intensity of the back-scattered light in the vicinity of the echo time. The $\Delta t$-integrated intensity is approximately proportional to the square of the $T$-dependent part of $\tilde{E}_2$ from \Eq \ref{eqn:E2-Theory}:
\begin{align}
\begin{split}
  \label{eqn:s2-Theory}
  \tilde{s}_2(T)
  & \sim \frac{u_2^4}{4} \sin^4(\omega_q (T + \Delta t) - \theta) \\
  & \times \left[ J_0(\kappa_2) + \frac{4}{3} J_2(\kappa_2) + \frac{1}{3} J_4(\kappa_2) \right]^2.
\end{split}
\end{align}
As a function $T$, this expression is periodic with fundamental frequency $2\omega_q$ (period $\tau_q = \pi/\omega_q \sim 32$ $\mu$s). The signal, $\tilde{s}_2$, is proportional to the grating contrast measured in the experiment, which is a manifestation of phase modulation in the atomic wave function following the interaction with two sw pulses. The response of the AI depends on the level structure of the atom and the coupling between these levels and the driving field. We now explore these effects in detail.

\subsection{Recoil signal including magnetic sub-levels}
\label{sec:Theory-mLevels}

So far, we have reviewed the theoretical expressions for the one and two-pulse recoil signals, including the effects of spontaneous emission, for a two-level atom. In the experiment we use $^{85}$Rb, which is a multi-level atom. If only the $F = 3 \to F' = 4$ transition is considered, there are $2F+1 = 7$ ground state magnetic sub-levels and 9 excited state sub-levels. These energetically degenerate sub-levels have a significant role on the response of the AI. A previous treatment of the two-pulse signal \cite{Kumar1} accounted for multiple atomic levels, but assumed the population was equally distributed and ignored effects due to spontaneous emission. Here, we extend the theoretical model of the recoil signal discussed in the previous two sub-sections to include multi-level atoms with an arbitrary distribution of initial sub-level populations.

The coupling strength between states $\ket{g} = \ket{n_g \, J_g \, m_g}$ and $\ket{e} = \ket{n_e \, J_e \, m_e}$ is determined by the dipole matrix element
\begin{align}
\begin{split}
  \label{eqn:mu_eg}
  \mu_{eg}
  & = - e\bra{e} \hat{\epsilon}_{q_L} \cdot \bm{r} \ket{g} \\
  & = - e \langle n_e \, J_e \| r \| n_g \, J_g \rangle C^{J_g\;\;1\;\;\;J_e}_{m_g\;q_L\;m_e},
\end{split}
\end{align}
where $n_g$, $n_e$ are the principal quantum numbers, $J_g$, $J_e$ are the total angular momenta and $m_g$, $m_e$ are the magnetic sub-levels of the ground and excited states, respectively. In our case, $J_g = F = 3$ and $J_e = F' = 4$. The unit vector $\hat{\epsilon}_{q_L}$ represents the polarization of the laser field. Linear and circular polarization states are denoted by $q_L$ ($q_L = 0$ for linear and $q_L = \pm 1$ for $\sigma^{\pm}$ polarizations, respectively). The factor $\langle n_e \, J_e \| r \| n_g \, J_g \rangle$ in \Eq \ref{eqn:mu_eg} is the reduced matrix element associated with the radial part of the wave functions---the magnitude of which is unimportant for this treatment and will be absorbed into the Rabi frequency, $\Omega_0 = \mu_{eg} E_0/\hbar$. The factor $C^{J_g\;\;1\;\;\;J_e}_{m_g\;q_L\;m_e}$ is the Clebsch-Gordan coefficient, which describes how strongly two states are coupled by the photon and depends on the particular transition. Since we are only concerned with electric dipole transitions, it is non-zero only for states that obey the selection rules: $J_e = J_g + 1$ and $m_e = m_g + q_L$.

\begin{figure}[!t]
  \subfigure{
    \label{fig:Recoil-Theory-mLevel-2Level-150ns}
    \includegraphics[width=0.23\textwidth]{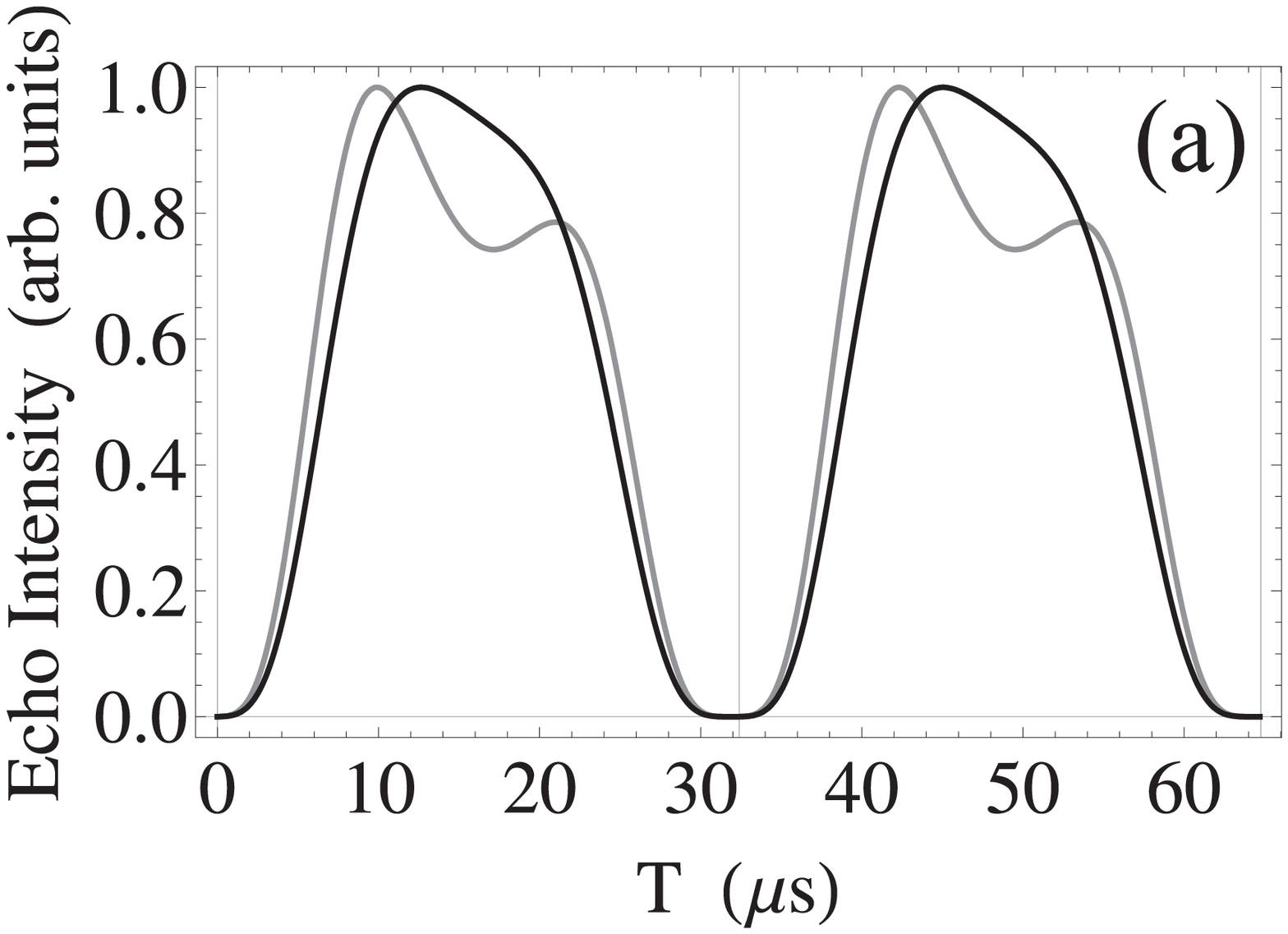}}
  \subfigure{
    \label{fig:Recoil-Theory-mLevel-2Level-200ns}
    \includegraphics[width=0.23\textwidth]{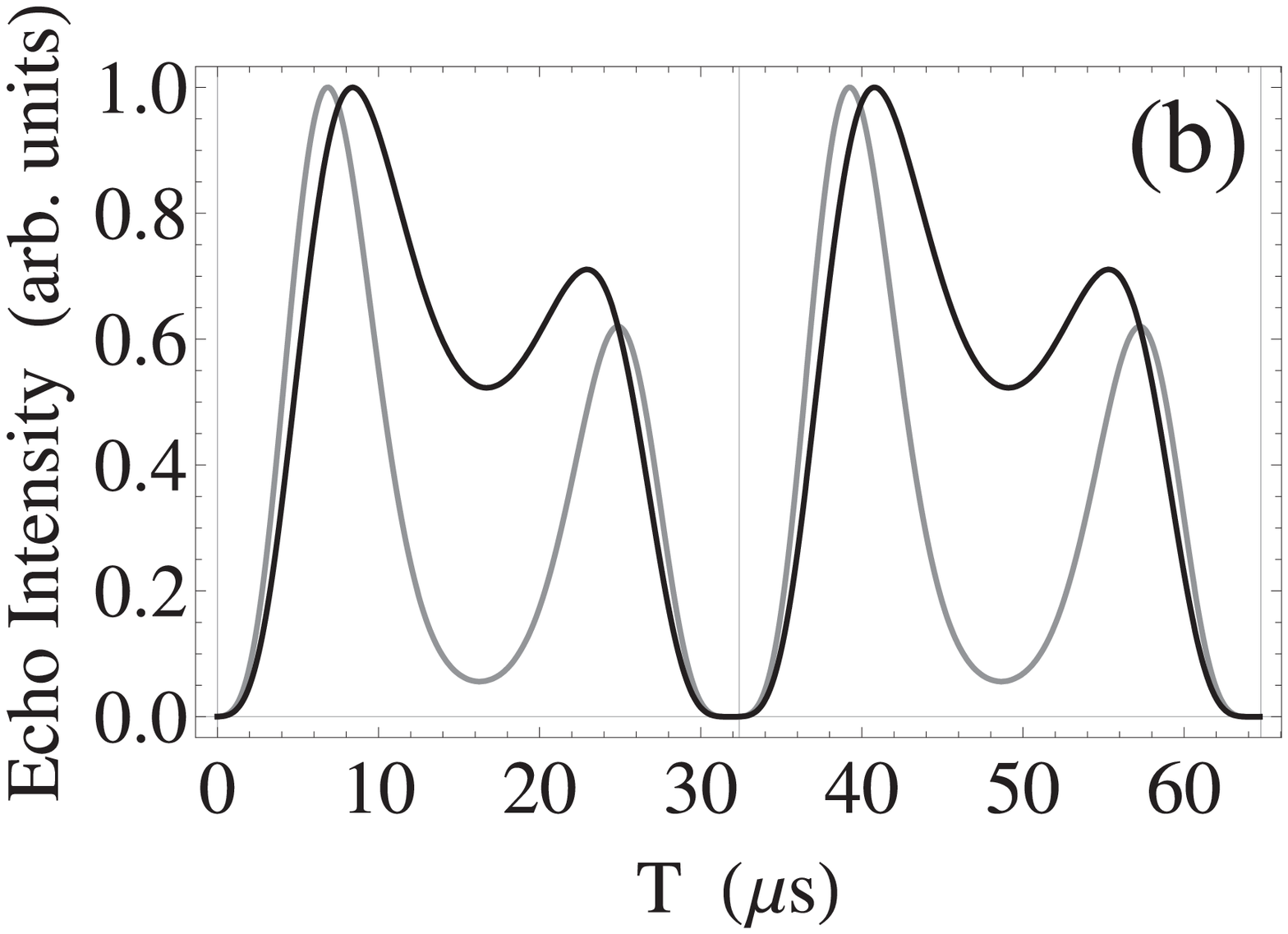}}
  \subfigure{
    \label{fig:Recoil-Theory-mLevel-2Level-250ns}
    \includegraphics[width=0.23\textwidth]{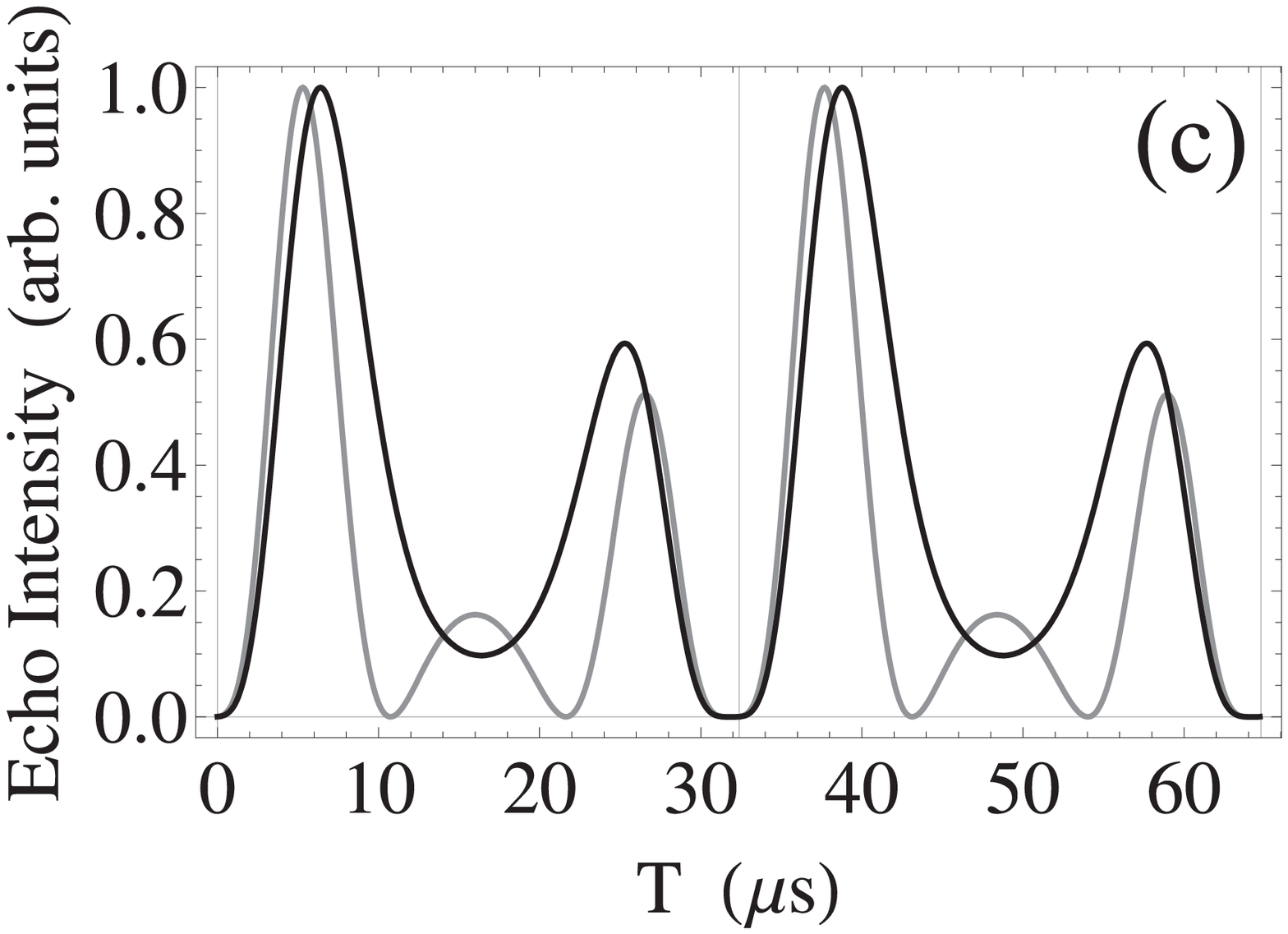}}
  \subfigure{
    \label{fig:Recoil-Theory-mLevel-2Level-300ns}
    \includegraphics[width=0.23\textwidth]{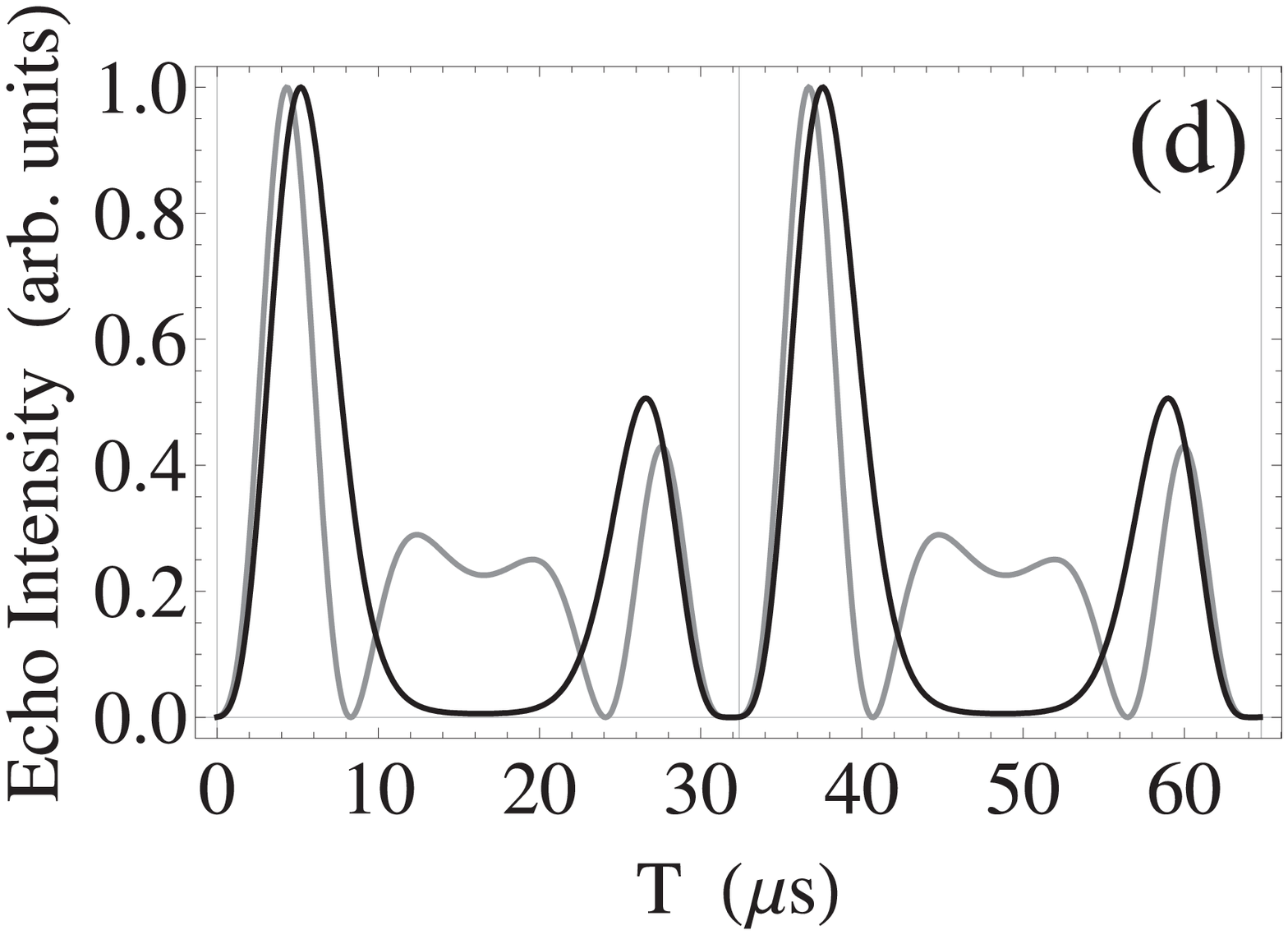}}
  \caption{Comparison of two-pulse recoil curves predicted by the two-level theory (\Eq \ref{eqn:s2-Theory}---gray curve) and the theory including magnetic sub-levels (square of \Eq \ref{eqn:E2-mg-sum}---black curve) for various $2^{\rm{nd}}$ pulse durations, $\tau_2$ (a: 150 ns; b: 200 ns; c: 250 ns; d: 300 ns). Here, the $m$-level populations were assumed to be equally distributed among the 7 levels of the $J_g = 3$ ground state. Pulse parameters: $\Delta = 10 \, \Gamma_{\rm{N}}$; $\Omega_0 = 2.5 \, \Gamma_{\rm{N}}$; $\Gamma = \Gamma_{\rm{N}}$.}
  \label{fig:Recoil-Theory-Comparison}
\end{figure}

From \Eq \ref{eqn:mu_eg}, it is apparent that each degenerate $m$-level interacts with a sw pulse (of a given polarization) with a different coupling strength---which is proportional to the Rabi frequency for each transition: $C^{J_g\;\;1\;\;\;J_e}_{m_g\;q_L\;m_e} \, \Omega_0$. In the experiment, this differential coupling causes the population of the $m$-levels to become unbalanced after the interaction with the sw pulse (optical pumping). The degree of the imbalance is determined by $\Omega_0$, $\Delta$ and the pulse durations. In the analytical treatment that follows, optical pumping is not taken into account. We assume the population of each $m$-level remains constant during the sw pulses. However, the numerical simulation to be discussed in \S\ref{sec:Sims} includes optical pumping effects.

The area of a given sw pulse, denoted by index $j = 1$ or 2, is given by
\be
  u_j^{(m_g)} = \frac{\Omega_0^2 \tau_j}{2\Delta} \left[ 1 + \left(\frac{\Gamma}{2\Delta} \right)^2 \right]^{-1/2} \left( C^{J_g\;\;1\;\;\;J_e}_{m_g\;q_L\;m_e} \right)^2
\ee
where $\tau_j$ is the pulse duration. For the one-pulse and two-pulse signals, the back-scattered field amplitude from the state $\ket{J_g\,m_g}$ is
\begin{subequations}
\label{eqn:Ej-mg-Theory}
\bea
  \label{eqn:E1-mg-Theory}
  \tilde{E}^{(m_g)}_1(t)
  & \propto & u^{(m_g)}_1 \left( C^{J_g\;\;1\;\;\;J_e}_{m_g\;q_L\;m_e} \right)^2 \\
  & \times  & \sin(\omega_q t - \theta) \left[ J_0 \left(\kappa_1\right) + J_2\left(\kappa_1\right) \right], \notag \\
  \label{eqn:E2-mg-Theory}
  \langle \tilde{E}_2^{(m_g)}(T) \rangle
  & \propto & \frac{1}{2} \big(u_2^{(m_g)}\big)^2 \left( C^{J_g\;\;1\;\;\;J_e}_{m_g\;q_L\;m_e} \right)^2 \notag \\
  & \times  & \sin^2 \big(\omega_q (T + \Delta t) - \theta\big) \\
  & \times  & \left[ J_0(\kappa_2) + \frac{4}{3} J_2(\kappa_2) + \frac{1}{3} J_4(\kappa_2) \right]. \notag
\eea
\end{subequations}
Here, it is understood that $u_1$ in \Eq \ref{eqn:kappa1} for $\kappa_1$ (for a two-level atom) has been replaced by $u_1^{(m_g)}$ for the multi-level case. Similarly, $u_2$ in \Eq \ref{eqn:kappa2} for $\kappa_2$ has been replaced by $u_2^{(m_g)}$. The extra factor of $\big(C^{J_g\;\;1\;\;\;J_e}_{m_g\;q_L\;m_e}\big)^2$ in \Eqs \ref{eqn:Ej-mg-Theory} arises due to the coupling of states $\ket{J_g\,m_g}$ and $\ket{J_e\,m_e}$ by the traveling wave read-out pulse (assuming that the scattered field has the same polarization as the read-out pulse). The total scattered field from the atom is proportional to the sum of the fields scattered by each $m$-level weighted by the final population of that level, $\Pi_{J_g\,m_g}$:
\begin{subequations}
\label{eqn:E-mg-sum}
\bea
  \label{eqn:E1-mg-sum}
  \tilde{E}_1(t) & \propto & \sum_{m_g} \Pi_{J_g\,m_g} \tilde{E}_1^{(m_g)}(t) \\
  \label{eqn:E2-mg-sum}
  \langle \tilde{E}_2(T) \rangle & \propto & \sum_{m_g} \Pi_{J_g\,m_g} \langle \tilde{E}_2^{(m_g)}(T) \rangle.
\eea
\end{subequations}
The one and two-pulse recoil signals are proportional to the square of the total back-scattered field amplitudes given by \Eqs \ref{eqn:E-mg-sum} ($\tilde{s}_j \propto |\tilde{E}_j|^2$, for index $j = 1,2$). The form of \Eqs \ref{eqn:E-mg-sum} allows for interference between scattered fields from each $m$-level. This additional interference from magnetic sub-levels strongly affects the shape of the recoil signal. Figure \ref{fig:Recoil-Theory-Comparison} shows a comparison of two-pulse recoil signals predicted by the two-level theory (\Eq \ref{eqn:s2-Theory}) and the theory including multiple sub-levels (square of \Eq \ref{eqn:E2-mg-sum}) for various second pulse durations, $\tau_2$. The two-level theory predicts extra zeroes in the signal shape that are not observed experimentally. For the same set of pulse parameters, the multi-level theory does not predict these extra zeroes due to the interference of back-scattered light from each magnetic sub-level. The multi-level theory models experimental data much more effectively than all previous models, as we will show in \S\ref{sec:Results-mLevels}.

\section{Description of simulation}
\label{sec:Sims}

The main goal of the simulation is to compute the two-pulse recoil signal (scattered field intensity as a function of pulse separation, $T$) under the conditions of the experiment. In practice, the signal in the experiment is obtained from a macroscopic sample (typically $\sim 10^9$ atoms) at a temperature of $\mathcal{T} \sim 100$ $\mu$K. The most basic assumption of the simulation is that the signal from the AI can, in principle, be produced by a single atom. Instead of velocity-averaging over many atoms with a well-defined momentum, as in the theoretical calculation of the two-pulse recoil signal, the simulation solves for the time-evolution of a single atom (represented by a wave-packet in momentum space) with a Maxwell-Boltzmann probability distribution of velocities corresponding to a given temperature.

The theoretical framework of the simulation uses the Schr\"{o}dinger picture to account for a number of physical effects, such as the motion of the atom at all times during the evolution; the bandwidth of sw pulses; Doppler shifts of momentum states; momentum state excitation for arbitrary sw pulse duration, field strength and detuning; population and coherence transfer between ground and excited states; population dissipation due to spontaneous emission; and optical pumping of magnetic sub-levels.

We solve for the time-dependent state amplitudes of the atomic wave function. The computational cost of this problem scales as $N_p$ multiplied by the number of time-steps used, where $N_p$ is the number of discrete momentum states. This is much less computationally expensive than the Heisenberg approach, where an $N_p \times N_p$ density matrix must be computed at every time-step. However, the Heisenberg approach allows for a much broader class of problems to be addressed, such as the atomic recoil due to spontaneous emission \cite{Berman1} or $N$-atom effects such as collective excitation and emission of radiation. Similarly, a Monte-Carlo wave function (MCWF) approach in the Schr\"{o}dinger picture would allow for a more complete model of spontaneous emission \cite{Molmer, Janicke1} with additional computational cost. In other work \cite{Beattie1}, we carried out MCWF simulations of sw pulses to study the excitation of momentum states and their relative populations for comparison with an experiment. In the case of the two-pulse recoil experiment, we found that the random photon recoil due to spontaneous emission has a negligible effect on the signal shape compared to the excited state population damping associated with spontaneous emission. As a result, the photon recoil associated with spontaneous emission is not taken into account in the treatment presented here.

We now review the theoretical framework of the simulation. We restrict our model to atoms with two manifolds of energetically degenerate magnetic sub-levels and total ground (excited) state angular momentum $J_g$ $(J_e = J_g + 1)$. We label the states of the atom as $\ket{J_g\;m_g}$ and $\ket{J_e\; m_e}$. The Schr\"{o}dinger equation is solved numerically in momentum space for wave functions in the field-interaction representation under the influence of an off-resonant, time-dependent standing wave potential. The Hamiltonian for this system in the rotating wave and dipole approximations is \cite{Berman1}
\begin{align}
\begin{split}
  \label{eqn:Hamiltonian}
  \mathcal{H}
  & = \frac{\hat{P}_z^2}{2M} + \frac{\hbar}{2} \left( \Delta - \frac{k P_z}{M} \right) (\hat{S}_{J_g} - \hat{S}_{J_e})
  + \frac{\hbar \Omega(t)}{2} \\
  & \times \big(\ket{p-\hbar k}\bra{p} \, + \, \ket{p+\hbar k}\bra{p} \big)
  \big(\hat{S}_{q_L}^+ + \hat{S}_{q_L}^-\big).
\end{split}
\end{align}
Here, $\hat{P}_z$ is the momentum operator along the $z$-axis; $\Delta \equiv \omega - \omega_0$ is the detuning of the laser field from resonance; $\hat{S}_{J_g} = \ket{J_g}\bra{J_g}$ $\big(\hat{S}_{J_e} = \ket{J_e}\bra{J_e}\big)$ is the ground (excited) state projection operator; $\Omega(t) = \mu E_0 \mathcal{E}(t)/\hbar$ is the (time-dependent) Rabi frequency with envelope function $\mathcal{E}(t)$; $\ket{p \pm \hbar k}\bra{p}$ are raising and lowering operators for the momentum state $\ket{p}$ in units of the photon momentum; and $\hat{S}_{q_L}^+$ $\big( \hat{S}_{q_L}^- \big)$ is a raising (lowering) operator proportional to the atomic dipole operator for laser field polarization state $q_L$. These operators are defined as follows
\begin{subequations}
\bea
  \hat{S}_{q_L}^+ \ket{J_g\; m_g}
  & = & C^{J_g\;\;1\;\;\;J_e}_{m_g\;q_L\;m_g + q_L} \ket{J_e\; m_g + q_L} \\
  \hat{S}_{q_L}^- \ket{J_e\; m_e}
  & = & C^{J_e\;\;1\;\;\;J_g}_{m_e\;q_L\;m_e - q_L} \ket{J_g\; m_e - q_L} \\
  \hat{S}_{q_L}^+ \ket{J_e\; m_e}
  & = & \hat{S}_{q_L}^- \ket{J_g\; m_g} = 0.
\eea
\end{subequations}
Clebsch-Gordan coefficients, $C^{j_1\;\;j_2\;\;\;J}_{m_1\;m_2\;M}$, describe the coupling of different ground and excited state $m$-levels. The pulse envelope, $\mathcal{E}(t)$, is based on a sum of exponential functions of the form $1/(1 + e^{-x})$, where $x = (t-t_0)/\tau_{\rm{rise}}$ for the rising edge of the pulse at time $t = t_0$, $x = -(t-t_1)/\tau_{\rm{rise}}$ for the falling edge at $t = t_1$ and $\tau_{\rm{rise}}$ is the rise time. Typically, the rise time is set equal to the experimentally measured value of 20 ns. The duration of the pulse is defined as $\tau = t_1 - t_0$.

The total momentum space wave function, $\Phi(p,t)$, can be written as a linear combination of the $2J_g+1$ ground states $\ket{J_g\;m_g}$ and the $2J_e+1$ excited states $\ket{J_e\;m_e}$:
\begin{align}
\begin{split}
 \Phi(p,t)
 & = \sum_{m_g} \alpha_{J_g\,m_g}(p,t) \ket{J_g\;m_g} \\
 & + \sum_{m_e}  \beta_{J_e\,m_e}(p,t) \ket{J_e\;m_e},
\end{split}
\end{align}
where $\alpha_{J_g\,m_g}$ and $\beta_{J_e\,m_e}$ are the corresponding state amplitudes.

The main goal of the simulation is to find the state amplitudes that simultaneously satisfy the Schr\"{o}dinger equation and the set of rate equations that describe spontaneous emission. The solution to the Schr\"{o}dinger equation is computed using a combination of two methods. When the interaction potential is turned off (i.e. $\mathcal{E}(t) = 0$) and the excited state population $\big( \Pi_{J_e}(t) \big)$ is zero, an analytical solution of the Schr\"{o}dinger equation in free space is used to compute $\Phi(p,t)$. At all other times, the solution is computed using a fourth order Runge-Kutta routine \cite{Burden}. To describe spontaneous transitions from the excited to ground levels, the numerical solver also simultaneously satisfies the following rate equations for the $m$-level populations:
\begin{subequations}
\bea
  \dot{\Pi}_{J_e\,m_e}
  & = & -\Gamma \, \Pi_{J_e\,m_e} \\
  \dot{\Pi}_{J_g\,m_g}
  & = & \sum_{q_L} \left(\, C^{J_g\;\;1\;\;\;J_e}_{m_g\;q_L\;m_g+q_L} \right)^2 \, \Gamma \,
  \Pi_{J_e\,m_g + q_L}.
\eea
\end{subequations}
These equations have a solution
\begin{subequations}
\bea
  \Pi_{J_e\,m_e}(t')
  & = & \Pi_{J_e\,m_e}(t) e^{-\Gamma (t' - t)} \\
  \Pi_{J_g\,m_g}(t')
  & = & \Pi_{J_g\,m_g}(t) + \Big( 1 - e^{-\Gamma(t' - t)} \Big) \\
  & \times & \sum_{q_L} \left(\, C^{J_g\;\;1\;\;\;J_e}_{m_g\;q_L\;m_g + q_L} \right)^2 \,
  \Pi_{J_e\,m_g + q_L}(t). \nonumber
\eea
\end{subequations}
Typically, $\Gamma = \Gamma_{\rm{N}} = 3.76 \times 10^7$ s$^{-1}$, which is the natural radiative rate of the $5^2$P$_{3/2}$ excited state for $^{85}$Rb \cite{Metcalf}.

\begin{figure}[!t]
  \subfigure{
    \label{fig:gPop-sq1}
    \includegraphics[width=0.38\textwidth]{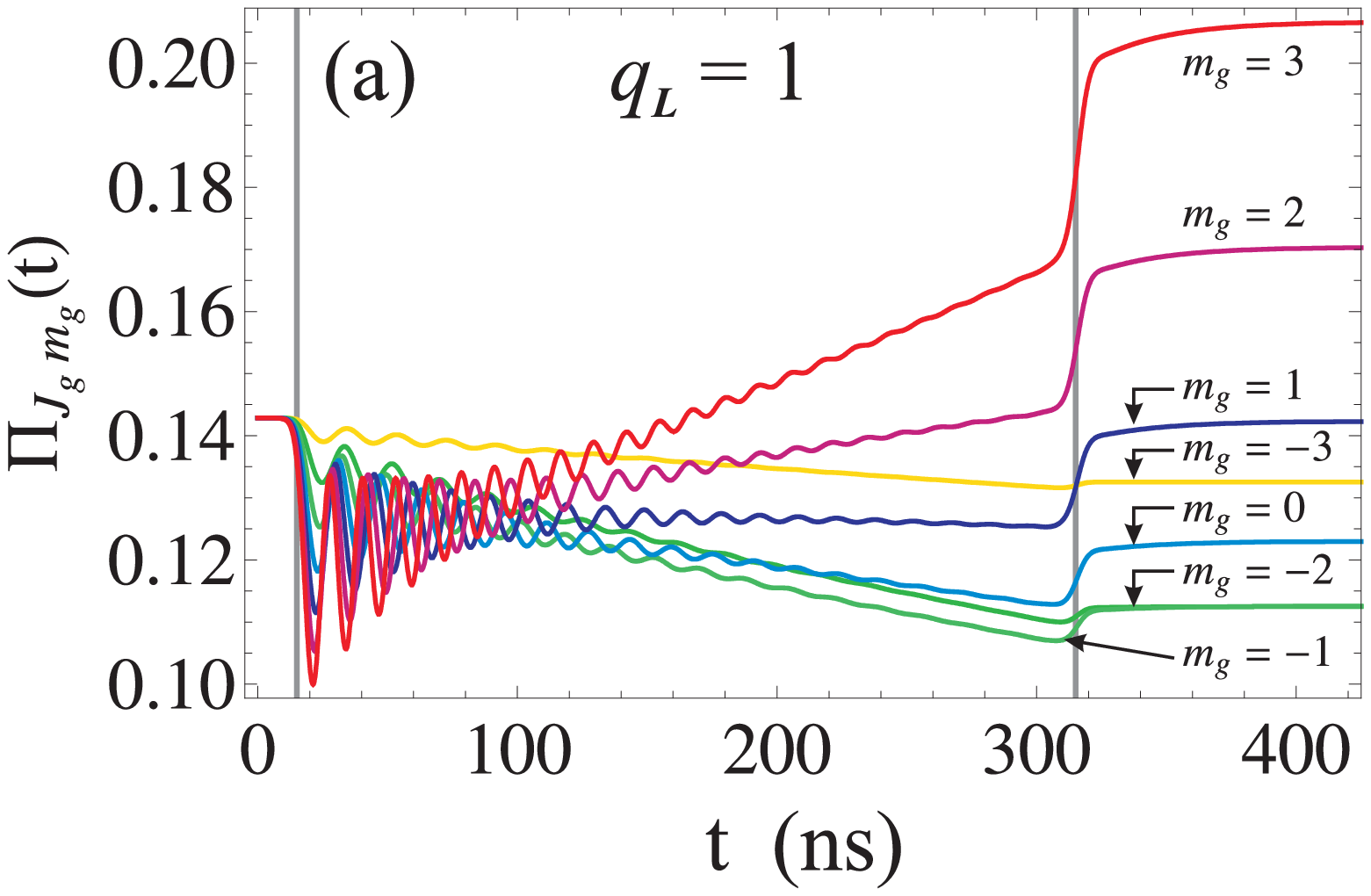}}
  \subfigure{
    \label{fig:gPop-sq0}
    \includegraphics[width=0.38\textwidth]{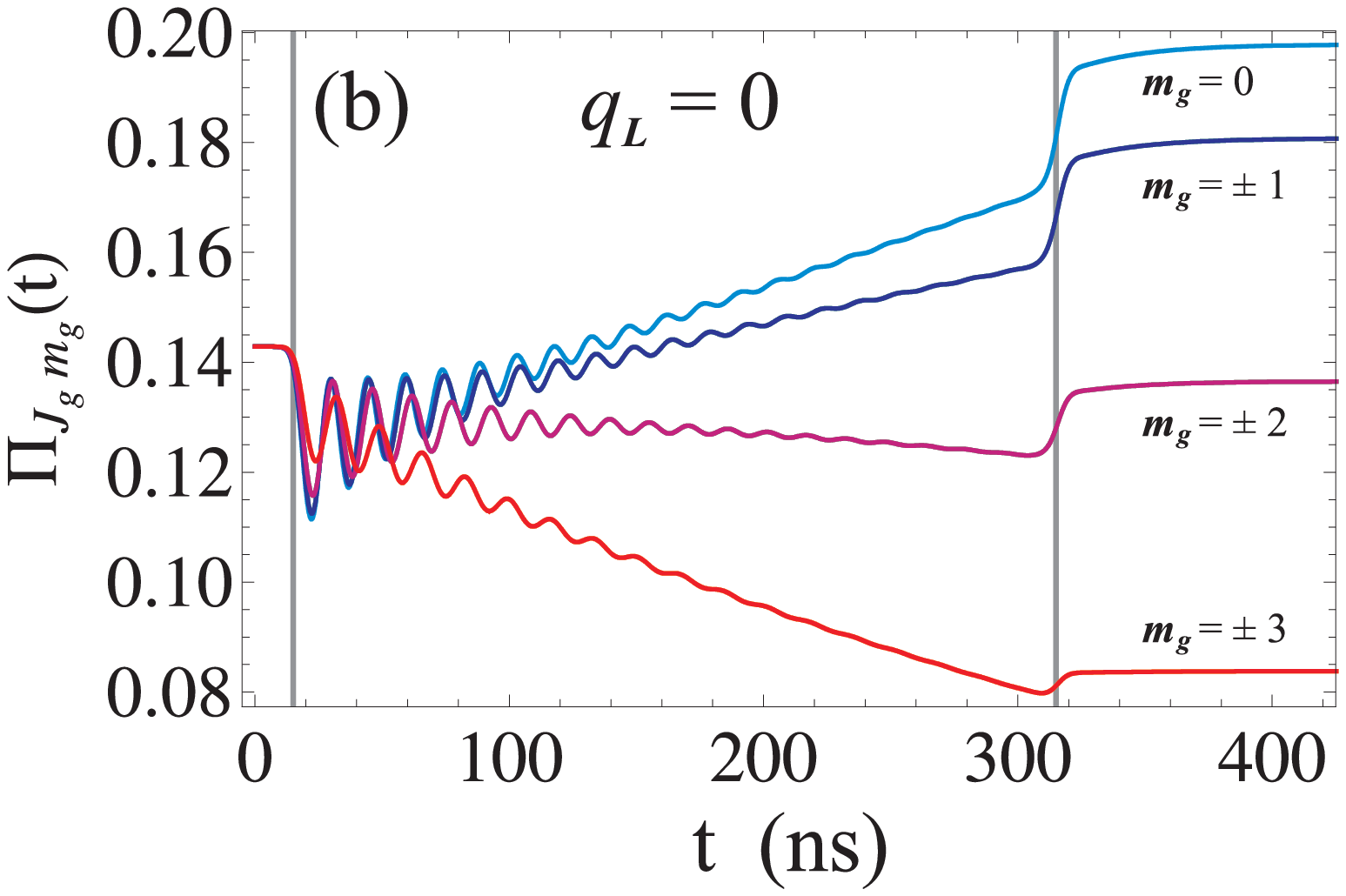}}
  \caption{(Color online) Evolution of ground state magnetic sub-level populations during a sw pulse for (a) $\sigma^{+}$ and (b) linear polarization states. The gray grid lines in each plot indicate where the sw pulse turns on and off. For circular polarization there is optical pumping towards the $m_g = 3$ level, while for linear polarization the population is pumped towards the $m_g = 0$ level. Pulse parameters: $\tau_1 = 300$ ns; $\tau_{\rm{rise}} = 20$ ns, $\Delta = 8.36 \, \Gamma_{\rm{N}}$; $\Omega_0 = 5 \, \Gamma_{\rm{N}}$; $\Gamma = \Gamma_{\rm{N}}$.}
  \label{fig:mPopEvol}
\end{figure}

Figure \ref{fig:mPopEvol} shows the evolution of $m$-level populations in the ground level during a sw pulse for (a) circular and (b) linear polarizations. Here, the initial populations are equally distributed among the 7 sub-levels. Evidence of optical pumping is clearly visible after only $\tau_1 \sim 300$ ns, which is the typical duration of sw pulses in the experiment.

The initial ground state population, $\Pi_{J_g}(0)$, is set to unity and the code ensures the normalization of the wave function is preserved for all times thereafter. The population of a particular manifold at any time, $\Pi_{J}(t)$, is the sum over all $m$-level populations, $\Pi_{J\,m}(t)$, in that manifold and is related to the state amplitude, $\alpha_{J\,m}(p,t)$, through the following expressions:
\begin{subequations}
\bea
  \label{eqn:PiJg(0)}
  \Pi_{J_g}(t) & = & \sum_{m_g} \Pi_{J_g\,m_g}(t) = 1 - \Pi_{J_e}(t), \\
  \label{eqn:PimgJg(0)}
  \Pi_{J_g\,m_g}(t) & = & \int |\alpha_{J_g\,m_g}(p,t)|^2 dp.
\eea
\end{subequations}
To account for the finite velocity distribution of the laser cooled sample, we use a Gaussian wave packet to describe the initial momentum distribution of each magnetic sub-level. The initial wave function can be written as
\begin{subequations}
\bea
  \Phi(p,0)
  & = & \sum_{m_g} \alpha_{J_g\,m_g}(p,0) \ket{J_g\;m_g}, \\
  \alpha_{J_g\,m_g}(p,0)
  & = & \alpha_{J_g\,m_g}(0) \left[ \frac{e^{-(p - p_0)^2/2 \sigma_p^2}}{\sqrt{2 \pi} \sigma_p} \right]^{1/2}.
\eea
\end{subequations}
Here, $p_0$ is the average initial momentum, $\sigma_p = \hbar/\Lambda$ is the standard deviation of the $p$-space probability distribution and $\Lambda = h/(2 \pi M k_B \mathcal{T})^{1/2}$ is the thermal de Broglie wavelength. The coefficients $\alpha_{J_g\,m_g}(0)$ are free parameters that define the initial population of each ground state $m$-level.

For the one-pulse recoil signal, the back-scattered field from each $m$-level is computed using the amplitude of the $q$-Fourier harmonic from each ground state amplitude, as given by
\be
\begin{split}
  \label{eqn:E1-mg-Sim}
  E_1^{(m_g)}(t)
  & \propto \left( C^{J_g\;\;1\;\;\;J_e}_{m_g\;q_L\;m_e} \right)^2 \\
  & \times \int \alpha_{J_g\,m_g}^{*}(p,t) \alpha_{J_g\,m_g}(p - \hbar q,t) dp.
\end{split}
\ee
This expression is analogous to \Eq \ref{eqn:E1-mg-Theory} in the analytical model. The total back-scattered field is a sum over the scattered fields from each $m$-level
\be
  \label{eqn:E1-Sim}
  E_1(t) = \sum_{m_g} E_1^{(m_g)}(t).
\ee
No $m$-level population factors appear in \Eq \ref{eqn:E1-Sim}, as they do in the analytical equivalent (\Eq \ref{eqn:E1-mg-sum}), because the populations are intrinsically built into the state amplitudes, $\alpha_{J_g\,m_g}$, that appear in \Eq \ref{eqn:E1-mg-Sim}. The one-pulse recoil signal is then $s_1(t) = |E_1(t)|^2$, where $t$ is the time after the sw pulse. From \Eq \ref{eqn:E1-Sim}, it is evident that the shape of the recoil signal depends on the distribution of initial ground state amplitudes, $\alpha_{J_g\,m_g}(0)$.

In the two-pulse regime we use a temperature of $\sim 50$ $\mu$K, which is 5000 times larger than that used in the one-pulse regime, but a factor of $\sim 2$ smaller than the typical temperature in the experiment. This is done to reduce the computation time in the two-pulse regime. We use a similar method to compute the intensity of the back-scattered field as in the two-pulse experiment. In the vicinity of the grating echo the intensity is $|E_1(2T + t')|^2$, where $t'$ is the time relative to $t = 2T$ and $E_1$ is given by \Eq \ref{eqn:E1-Sim}. Therefore, we calculate the two-pulse recoil signal by integrating this quantity over $t'$, as given by
\be
  \label{eqn:s2-Sim}
  s_2(T) = \int |E_1(2T + t')|^2 dt'.
\ee
This quantity is analogous to the expression for the two-pulse recoil signal for a two-level system (\Eq \ref{eqn:s2-Theory}) or for an $m$-level system (square of \Eq \ref{eqn:E2-mg-sum}).

To conclude the description of the simulations, we discuss the discretization of momentum space. In the two-pulse regime, the width of the wave function is $\sigma_p \sim 5 \hbar k$. The discretization size is then typically set to $\Delta p \sim \sigma_p/250 \sim \hbar k/50$ for good $p$-space resolution. However, due to the reduced temperature in the one-pulse regime, the width of the wave function is much smaller ($\sigma_p \sim 0.01 \hbar k$---much less than the $\hbar k$ transferred by absorption or emission along $z$). Hence, we can use $\Delta p = \hbar k$ since this is the smallest momentum transfer allowed by the sw interaction.

The number of sw pulses and the $p$-space discretization size has a large effect on the computational cost of the simulation in the two regimes. Since the two-pulse regime requires an additional sw pulse, a smaller $\Delta p$, and the evaluation of the signal is carried out as a function of $T$, it is much more computationally demanding than the one-pulse regime. Typically, the computation time is $\sim 12$ hours for the two-pulse regime, and $\sim 30$ seconds for the one-pulse regime on a computing network using 2.2 GHz processors (SHARCNET). However, the underlying physics in both regimes is essentially the same and many features of the two-pulse recoil signal can be studied using one-pulse simulations.

\section{Results and Discussion}
\label{sec:Results}

Here we discuss the two main results of this work: the effects on the recoil signal due to magnetic sub-levels and the momentum-dependent phase imprinted on the ground state by spontaneous and stimulated processes. We also present, for the first time, a detailed description of the mechanisms that contribute to the shape of the signal.

\subsection{Effects due to magnetic sub-levels}
\label{sec:Results-mLevels}

\begin{figure}[!t]
  \subfigure{
    \label{fig:Data-Theory-mLevels-70ns}
    \includegraphics[width=0.38\textwidth]{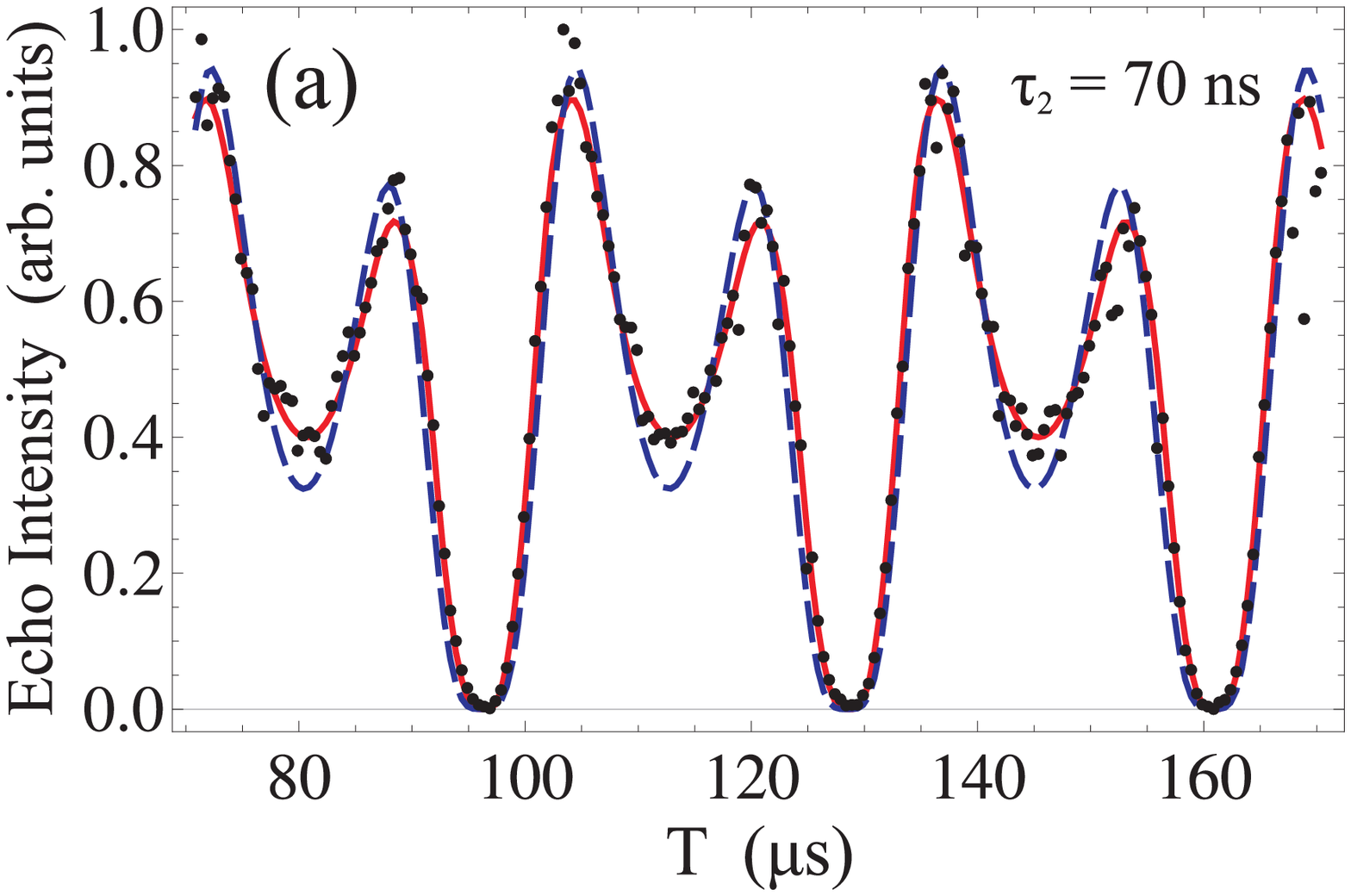}}
  \subfigure{
    \label{fig:Data-Theory-mLevels-86ns}
    \includegraphics[width=0.38\textwidth]{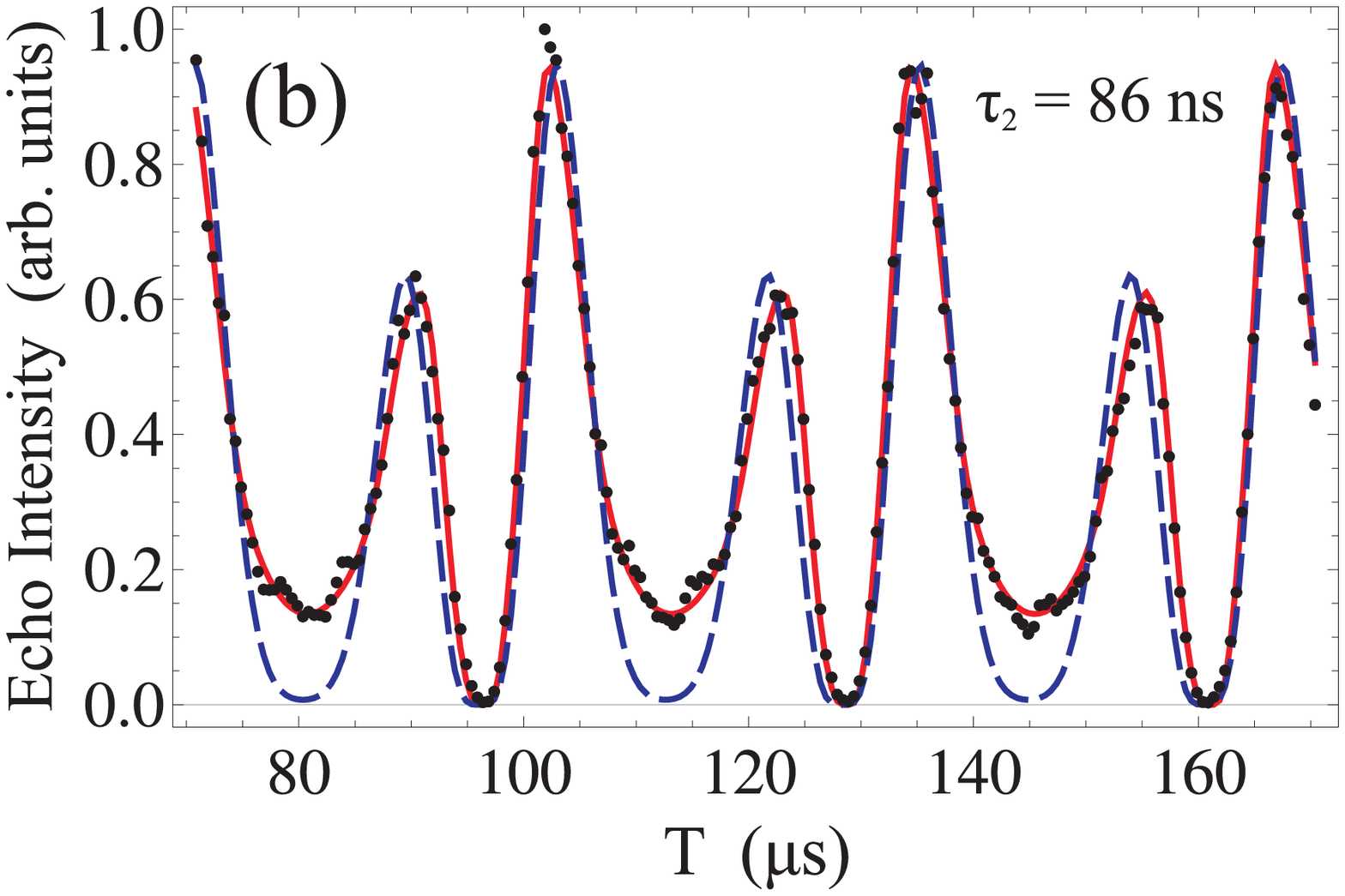}}
  \subfigure{
    \label{fig:Data-Theory-mLevels-98ns}
    \includegraphics[width=0.38\textwidth]{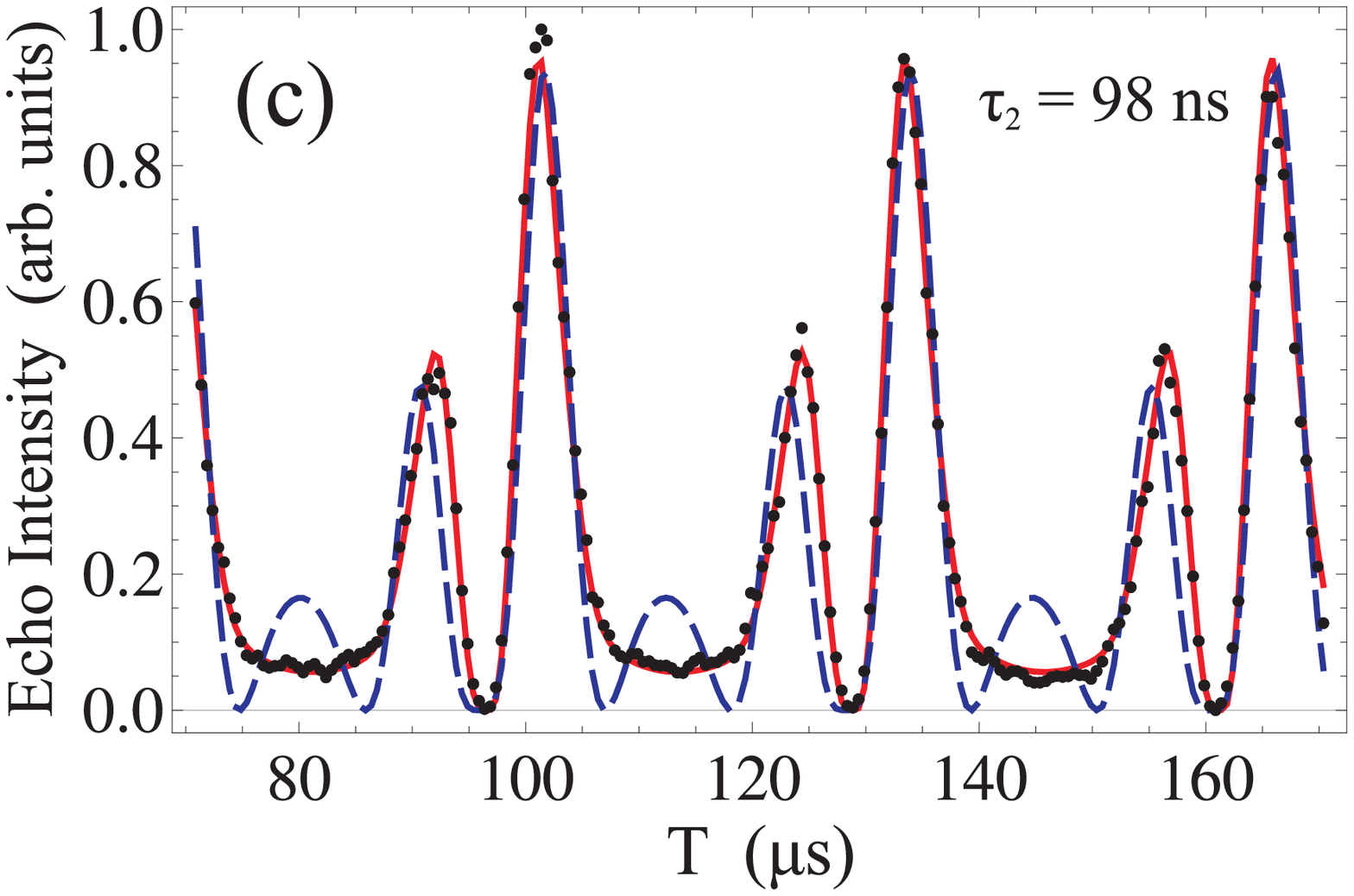}}
  \caption{(Color online) Data from the two-pulse recoil experiment for different 2$^{\rm{nd}}$ pulse durations, $\tau_2$ (a: 70 ns; b: 86 ns; c: 98 ns). Data is fit to the two-level expression (\Eq \ref{eqn:s2-Theory}) shown as the blue dashed line, resulting in a $\chi^2/\rm{dof} \sim 9 \times 10^{-3}$, where dof is the number of degrees of freedom. The data is also fit to the multi-level expression (square of \Eq \ref{eqn:E2-mg-sum}) shown as the red solid line, resulting in $\chi^2/\rm{dof} \sim 8 \times 10^{-4}$. The $\chi^2$ is computed assuming equally weighted data points (uncertainty for each point equal to unity). Final $m$-level populations, $\Pi_{J_g\,m_g}$, were estimated from the multi-level fits and are tabulated in Table \ref{tab:Pis}. Pulse parameters: detuning $\Delta \sim 50$ MHz; intensity $I \sim 50$ mW/cm$^2$; polarization state $|q_L| = 1$; $1^{\rm{st}}$ pulse duration $\tau_1 = 300$ ns.}
  \label{fig:Data-Theory-mLevels}
\end{figure}

Figure \ref{fig:Data-Theory-mLevels} shows data from the two-pulse recoil experiment for various second sw pulse durations, $\tau_2$. The echo intensity was recorded as a function of the pulse separation, $T$, for circularly polarized sw pulses ($|q_L| = 1$). Particular values of $\tau_2$ (70, 86 and 98 ns) were chosen to illustrate a range of signal shapes typically observed in the experiment. Two separate fits to the data are shown, one based on the two-level model (\Eq \ref{eqn:s2-Theory}) and the other on the multi-level model (square of \Eq \ref{eqn:E2-mg-sum}). The two-level theory is insufficient to accurately model experimental data, whereas the multi-level theory fits all aspects of the signal for a large range of shapes. In particular, the multi-level theory successfully models the asymmetry and the broad valleys between zeroes of the signal that occur as the area of the second pulse is increased, as shown in \Fig \ref{fig:Data-Theory-mLevels-98ns}. Fits using the multi-level theory show a factor of $\sim 10$ improvement in the $\chi^2/\rm{dof}$ compared to that of the two-level theory, which corresponds to a factor of $\sim 3$ improvement in the relative uncertainty of the recoil frequency. Thus, the multi-level model is ideally suited for precision measurements of $\omega_q$ using this technique.

\begin{table}[!b]
  \begin{tabular}{cccccccc}
  \hline 
  $\tau_2$ [ns] & $\Pi_{\,3\,-3}$ & $\Pi_{\,3\,-2}$ & $\Pi_{\,3\,-1}$ & $\Pi_{\,3\,0}$
                & $\Pi_{\,3\, 1}$ & $\Pi_{\,3\, 2}$ & $\Pi_{\,3\, 3}$ \\
  \hline 
  70 & 0.00 & 0.03 & 0.32 & 0.48 & 0.15 & 0.02 & 0.00 \\
  86 & 0.00 & 0.03 & 0.38 & 0.32 & 0.21 & 0.05 & 0.01 \\
  98 & 0.00 & 0.00 & 0.50 & 0.32 & 0.15 & 0.02 & 0.01 \\
  \hline 
  \end{tabular}
  \caption{Final $m$-level population estimates from multi-level fits to experimental data for various $2^{\rm{nd}}$ pulse durations, as shown in \Fig \ref{fig:Data-Theory-mLevels}. The fits were performed assuming $q_L = 1$.}
  \label{tab:Pis}
\end{table}

The development of this multi-level model---which contains only measurable parameters---represents a significant improvement in our understanding of the AI since previous efforts \cite{Beattie3} relied upon a phenomenological model to fit experimental data. This model was unable to explain the underlying physical mechanisms that govern the signal shape. Additionally, the present model is far more successful in fitting the full range of signal shapes that can be generated in the experiment compared to the phenomenological model.

Another complication in modeling the AI is that magnetic sub-levels and the spatial intensity profile of the excitation beams can be shown to produce similar effects on the signal shape. In this work, the beam diameter ($\sim 2$ cm) was larger than the diameter of the atomic cloud ($\sim 1$ cm). Therefore, we were able to demonstrate conclusively that magnetic sub-levels played the dominant role on the response of the AI.

The data shown in \Fig \ref{fig:Data-Theory-mLevels} validates the multi-level model developed in \S\ref{sec:Theory-mLevels}. A distribution of populations among several ground state magnetic sub-levels smears out any extra zeroes in the signal that would occur if the system were optically pumped into a single state, for example $\ket{J_g\,J_g}$. This smearing is due to interference between the coherently scattered light from each $m$-level. Additionally, the valleys between the zeroes in the signal are significantly broadened. These effects are most prominent when the pulse areas ($u_1$ and $u_2$) are large, since (1) higher order momentum states are contributing to the signal from each $m$-level (resulting in the double-peaked shape) and (2) population imbalance in the $m$-levels is maximized.

It is also possible to estimate the final $m$-level populations, $\Pi_{J_g\,m_g}$, from the multi-level fits to the data in \Fig \ref{fig:Data-Theory-mLevels}. Although we found the statistical errors in the populations from the fits to be relatively large, the estimates are considered accurate for two reasons. Firstly, a variation between the $\Pi_{J_g\,m_g}$ on the order of $\sim10$\% results in a deviation in the minimum of the $\chi^2$. Secondly, the distribution of populations is similar to those inferred from simulations. We attribute the large statistical errors to the presence of 12 free parameters in the fit function (based on the square of \Eq \ref{eqn:E1-mg-sum}), namely $u_2$, $\theta$, $\Delta t$, $\omega_q$, an amplitude factor $A$, and the 7 populations, $\Pi_{J_g\,m_g}$. From Table \ref{tab:Pis}, the general trend in the final populations is to move towards the $m_g > 0$ levels as the pulse duration increases. This is consistent with our expectations for $\sigma^+$ excitation pulses. Magnetic sub-level populations can be confirmed using independent techniques \cite{Fatemi}.

\begin{figure}[!t]
  \subfigure{
    \label{fig:q01Data-150ns}
    \includegraphics[width=0.23\textwidth]{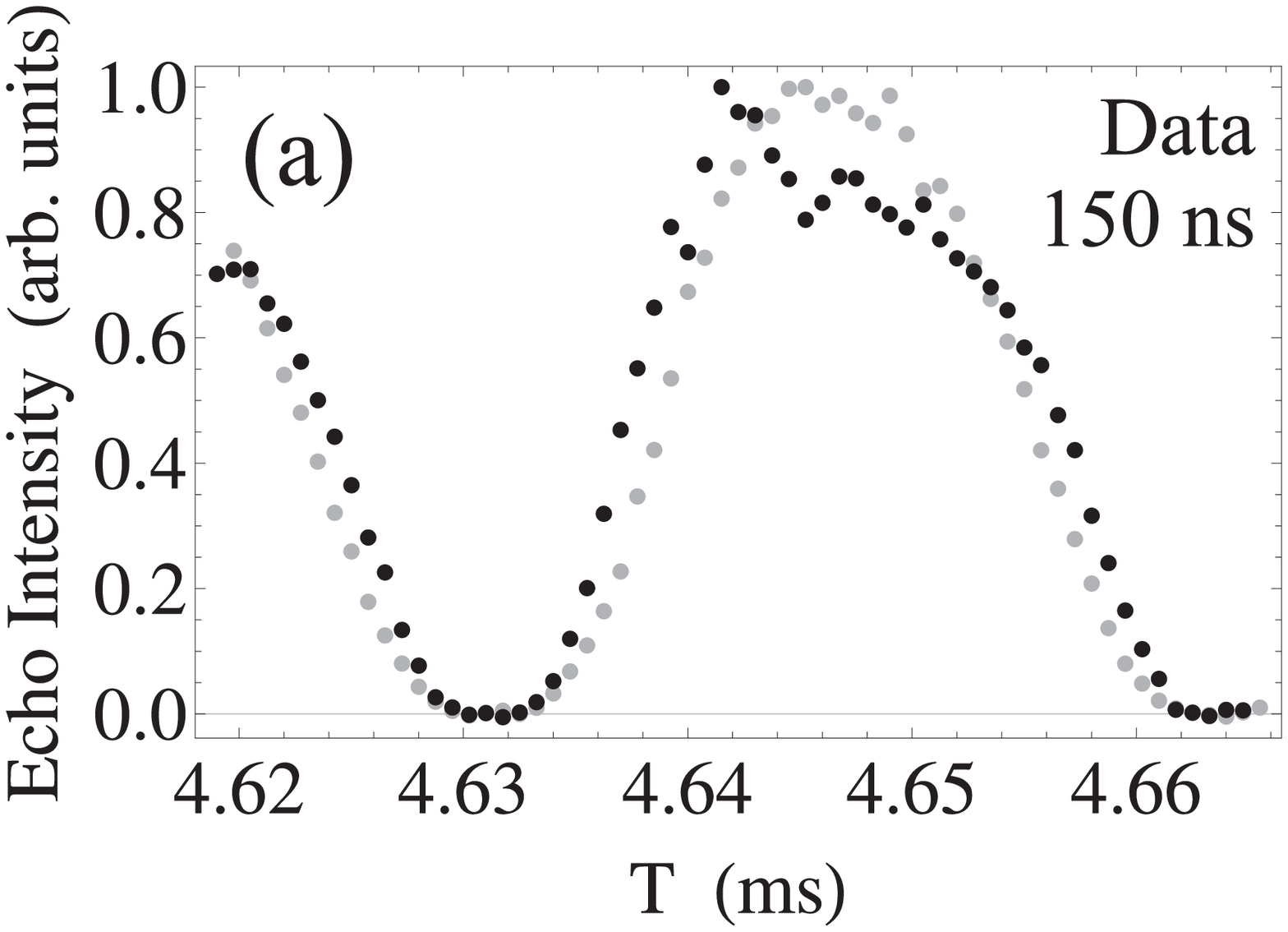}}
  \subfigure{
    \label{fig:q01Sims-150ns}
    \includegraphics[width=0.23\textwidth]{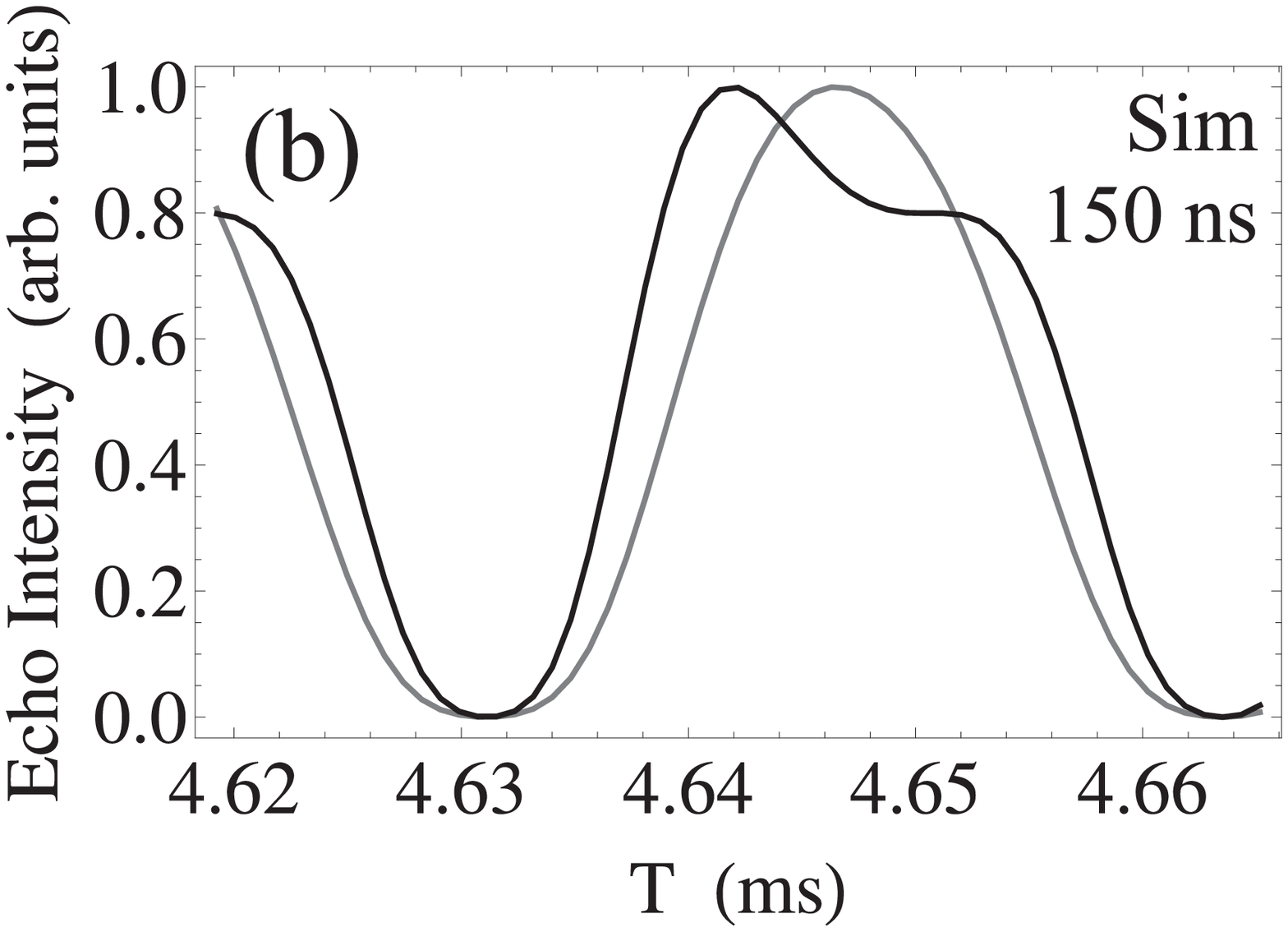}}
  \subfigure{
    \label{fig:q01Data-250ns}
    \includegraphics[width=0.23\textwidth]{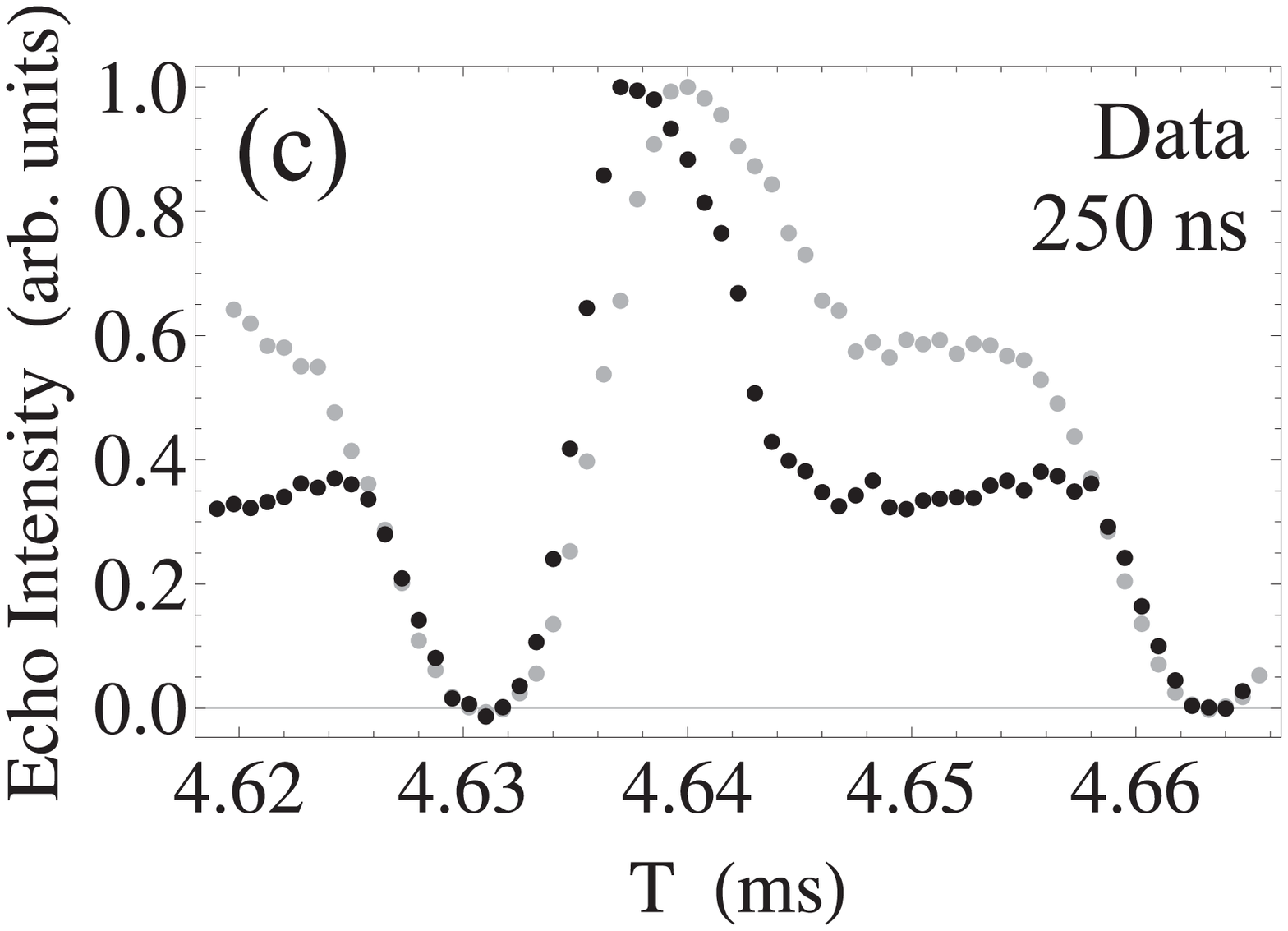}}
  \subfigure{
    \label{fig:q01Sims-250ns}
    \includegraphics[width=0.23\textwidth]{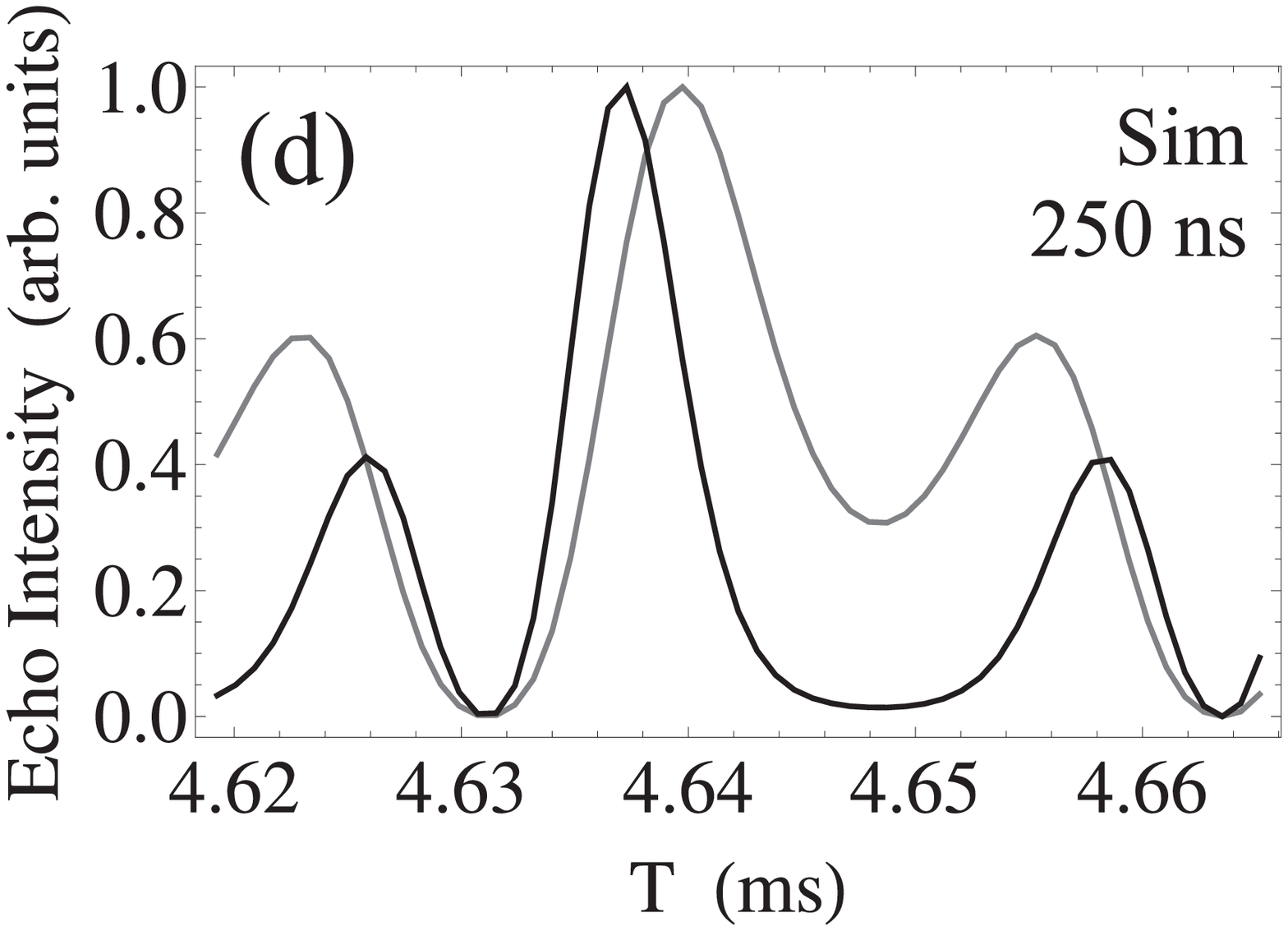}}
  \caption{Data from the two-pulse recoil experiment (a, c) and corresponding simulations (b, d) for two laser polarization states (gray: $q_L=0$; black: $|q_L|=1$) and two $2^{\rm{nd}}$ pulse durations (a, c: $\tau_2 = 150$ ns; b, d: $\tau_2 = 250$ ns). The recoil signals are qualitatively different for the two polarizations because of effects due to magnetic sub-levels. In the simulation, we used initial $m$-level populations $\Pi_{J_g\,m_g}(0) = \{0.05, 0.10, 0.20, 0.30, 0.20, 0.10, 0.05 \}$, where $m_g = -3$ on the left and $m_g = 3$ on the right, which produced the best agreement with the data for measured pulse parameters. Pulse parameters: detuning $\Delta \sim 40$ MHz; intensity $I \sim 13$ mW/cm$^2$; $1^{\rm{st}}$ pulse duration $\tau_1 = 500$ ns; pulse rise time $\tau_{\rm{rise}} = 20$ ns.}
  \label{fig:q01Data}
\end{figure}

Figure \ref{fig:q01Data} shows data from the two-pulse recoil experiment obtained using circularly polarized beams, and separately for linearly polarized beams. This data provides alternate confirmation of the role of magnetic sub-levels.
The recoil curves for each polarization state are qualitatively different for various second pulse durations. This figure also shows simulations that qualitatively agree with the data.

The signal shapes shown in \Fig \ref{fig:q01Data} for each polarization are consistent with our expectations based on two considerations. Firstly, the back-scattered field amplitude from each $m$-level is proportional to $\big( C^{J_g\;\;1\;\;\;J_e}_{m_g\;q_L\;m_e} \big)^2 \Pi_{J_g\,m_g}$. For the same distribution of $m$-level populations at the time of the echo, the two polarizations of the read-out pulse induce different atomic responses and therefore produce distinct signal shapes for the two cases. Secondly, the atom-field couplings are larger toward the $m_g = J_g$ state for the $q_L = 1$ polarization state ($C^{3\;1\;4}_{3\;1\;4} = 1$) compared to the couplings for the $m_g = 0$ state ($C^{3\;1\;4}_{0\;1\;1} \sim 0.598$). As a result, in the presence of spontaneous emission, circularly polarized sw pulses produces extreme state pumping. Similarly, for a linearly polarized sw pulse, the atom-field couplings are largest for the $m_g = 0$ state ($C^{3\;1\;4}_{0\;0\;0} \sim 0.756$, while $C^{3\;1\;4}_{3\;0\;3} = 0.5$). Consequently, there is optical pumping toward the $m_g = 0$ state. In this case, however, the atom-field coupling is not as strong as it is for the $m_g = J_g$ state with a $\sigma^+$-polarized beam. Therefore, for a given field strength, higher order momentum states will be populated for circularly polarized excitations. As a result of these two considerations, the recoil curves are qualitatively different for linear and circular polarizations.

\begin{figure}[!t]
  \subfigure{
    \label{fig:Feb24-SimDataOverlay-58ns}
    \includegraphics[width=0.38\textwidth]{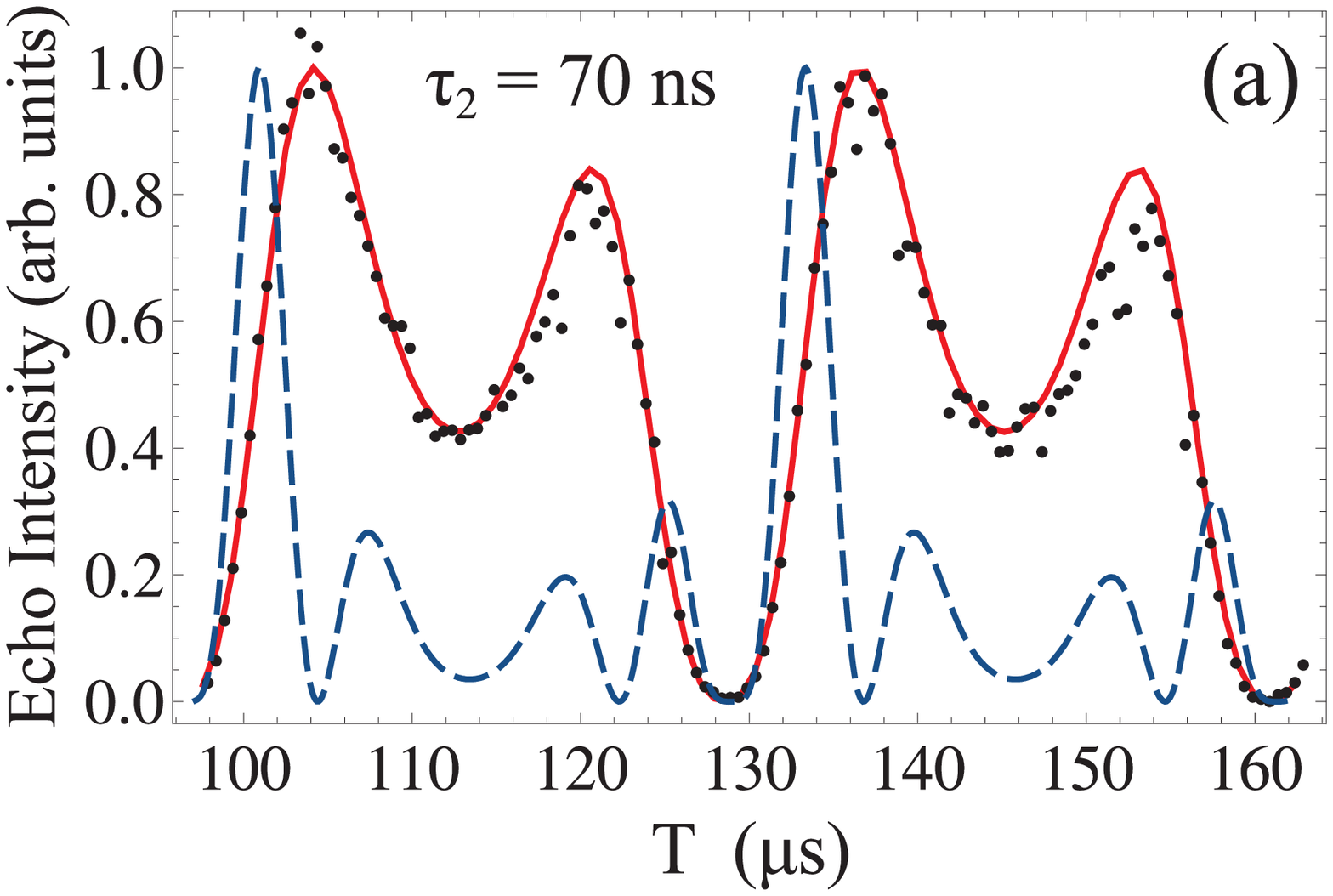}}
  \subfigure{
    \label{fig:Feb24-SimDataOverlay-98ns}
    \includegraphics[width=0.38\textwidth]{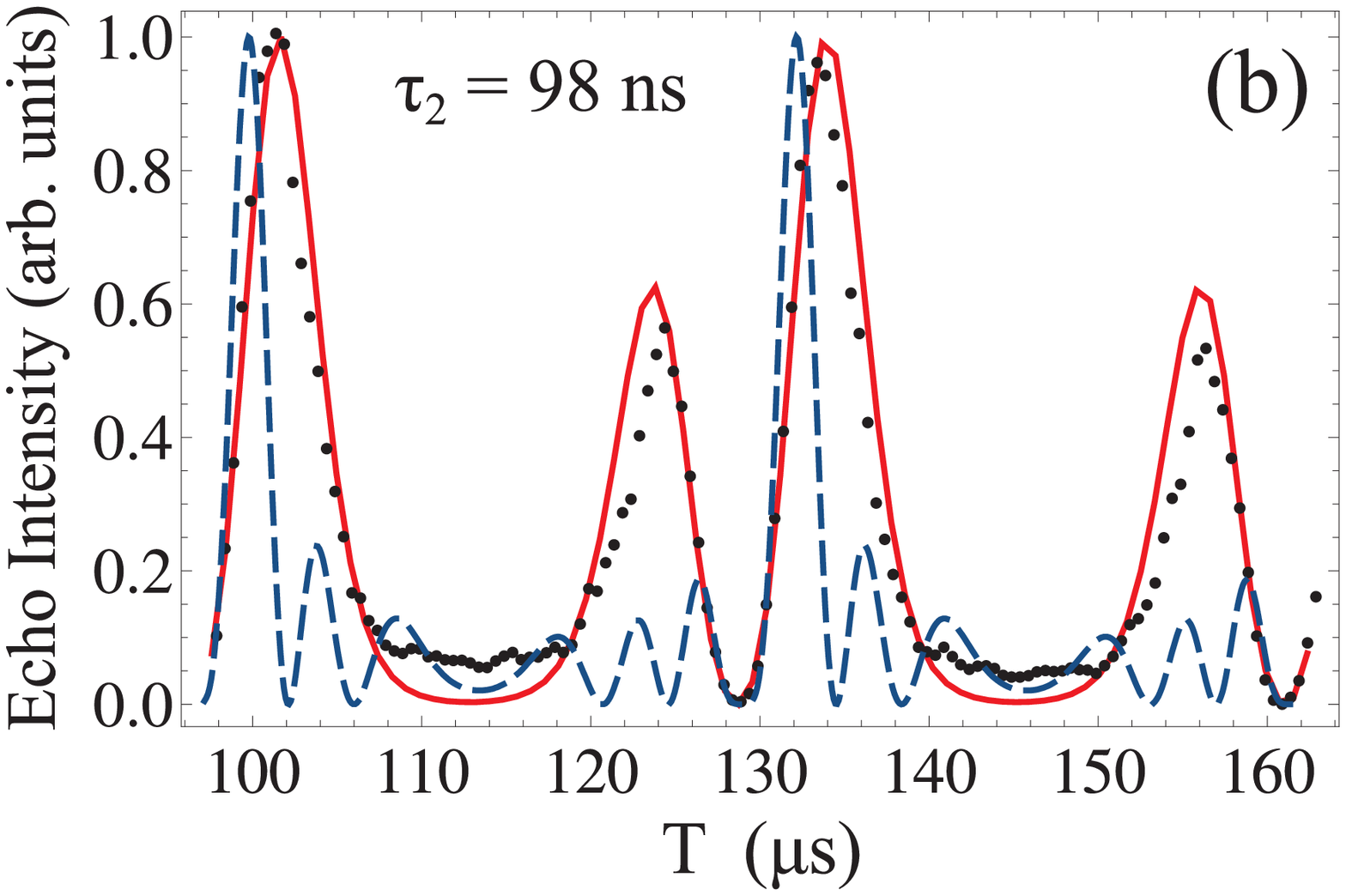}}
  \caption{(Color online) Data from the two-pulse recoil experiment for 2$^{\rm{nd}}$ pulse durations $\tau_2 = 70$ ns (a) and 98 ns (b) overlayed with predictions of the two-level theoretical model (\Eq \ref{eqn:s2-Theory}---dashed blue curve) and a multi-level simulation (solid red curve). No fits were performed here. There is good agreement between the simulation and the data for measured pulse parameters. In contrast, the two-level theory gives a poor prediction of the data for the same parameters. The distribution of initial $m$-level populations in the simulations was set to be similar to those shown Table \ref{tab:Pis}: $\Pi_{J_g\,m_g}(0) = \{ 0.05, 0.10, 0.25, 0.35, 0.20, 0.03, 0.02 \}$. Pulse parameters: $\tau_1 = 300$ ns; $\tau_{\rm{rise}} = 20$ ns; $\Delta \sim 8.36 \, \Gamma_{\rm{N}}$; $\Omega_0 \sim 5.2 \, \Gamma_{\rm{N}}$.}
  \label{fig:Feb24-SimDataOverlay}
\end{figure}

Figure \ref{fig:Feb24-SimDataOverlay} shows data overlayed with the prediction of the two-level model (\Eq \ref{eqn:s2-Theory}) as well as output from numerical simulations for the conditions of the experiment. Only a temporal shift and amplitude scaling has been applied to the predictions of the analytical model and the simulation---no fits were performed in this plot. It is clear that the simulation successfully models the data for measured pulse parameters such as the detuning, pulse duration and intensity. The initial sub-level populations are not measured in the experiment and were assumed to be similar to those measured in Table \ref{tab:Pis} (distribution peaked near $m_g = 0$). In contrast, the two-level model gives a poor prediction of the signal shape when the measured pulse parameters are used as inputs. These results suggest that the simulation can be used to study a variety of properties associated with this interferometer, including possible systematics that would affect a precision measurement of the recoil frequency. We discuss other applications of the simulation in \S\ref{sec:Conclusion}.

\subsection{Phase evolution during standing wave interaction}

We now describe the phase evolution of momentum states subject to standing wave excitation. We separate the discussion into three parts: phase dynamics due to spontaneous emission; phase dynamics due to stimulated emission and absorption; and a measurement of the combined effects including magnetic sub-levels. These effects are expected to be important for matter wave interference when the excited state contributes significantly to the relative phase between the different momentum states of the ground level. We describe these effects in the context of a single sw interaction using a theoretical treatment and compare the theoretical predictions to the results of simulations. Finally, we observe trends associated with these effects using the echo AI and verify these observations using two-pulse simulations.

\subsubsection{Spontaneous emission}
\label{sec:Results-SE}

In this section, we summarize our theoretical understanding of the influence of spontaneous emission on the shape of the recoil signal. The expression for the one-pulse recoil signal (\Eq \ref{eqn:s1-Theory}) contains the essential physics relevant for this discussion. The consequences of spontaneous emission on the two-pulse recoil signal (\Eq \ref{eqn:s2-Theory}) stem from the same effects as those in the one-pulse signal.

\begin{figure}[!t]
  \includegraphics[width=0.38\textwidth]{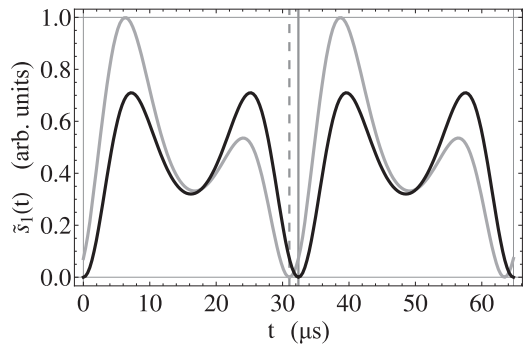}
  \caption{Comparison of recoil curves (back-scattered field intensity as a function of time) predicted by the one-pulse theory (\Eq \ref{eqn:s1-Theory}) with and without spontaneous emission. The black curve corresponds to a spontaneous emission-free system ($\Gamma = 0 \to \theta = 0$), while the gray curve corresponds to a system where spontaneous emission is present ($\Gamma = \Gamma_{\rm{N}} \to \theta = -0.133$ rad). The solid grid line shows the first zero in the signal after $t = 0$ in the absence of spontaneous emission, while the dashed grid line shows the zeroes shift by a temporal amount $\delta t = \theta/\omega_q$ in the presence of spontaneous emission. Pulse parameters: $\Delta = 7.5 \, \Gamma_{\rm{N}}$; $\Omega_0 = 1.5 \, \Gamma_{\rm{N}}$; $\tau_1 = 250$ ns; pulse area $u_1 \sim 1.43$.}
  \label{fig:SE-Effects}
\end{figure}

There are two main features that are observed as a consequence of including spontaneous emission in the theory. Firstly, within a single period, maxima adjacent to the signal zeroes have different amplitudes. We refer to this as the asymmetry in the recoil signal. Secondly, there is a temporal shift of the zeroes in the signal toward earlier times relative to the zeroes without spontaneous emission. Both of these features are best demonstrated when the pulse area is large---causing the double-peak structure within each period of the recoil signal. This is illustrated in \Fig \ref{fig:SE-Effects} on the basis of \Eq \ref{eqn:s1-Theory} and has been observed experimentally (see \Figs \ref{fig:Data-Theory-mLevels}, \ref{fig:q01Data} and \ref{fig:Feb24-SimDataOverlay}). Both of these features are present only when the spontaneous emission parameter, $\theta$, is non-zero (see \Eq \ref{eqn:theta}).

\begin{figure}[!t]
  \subfigure{
    \label{fig:ReE1-Components-NoSE}
    \includegraphics[width=0.38\textwidth]{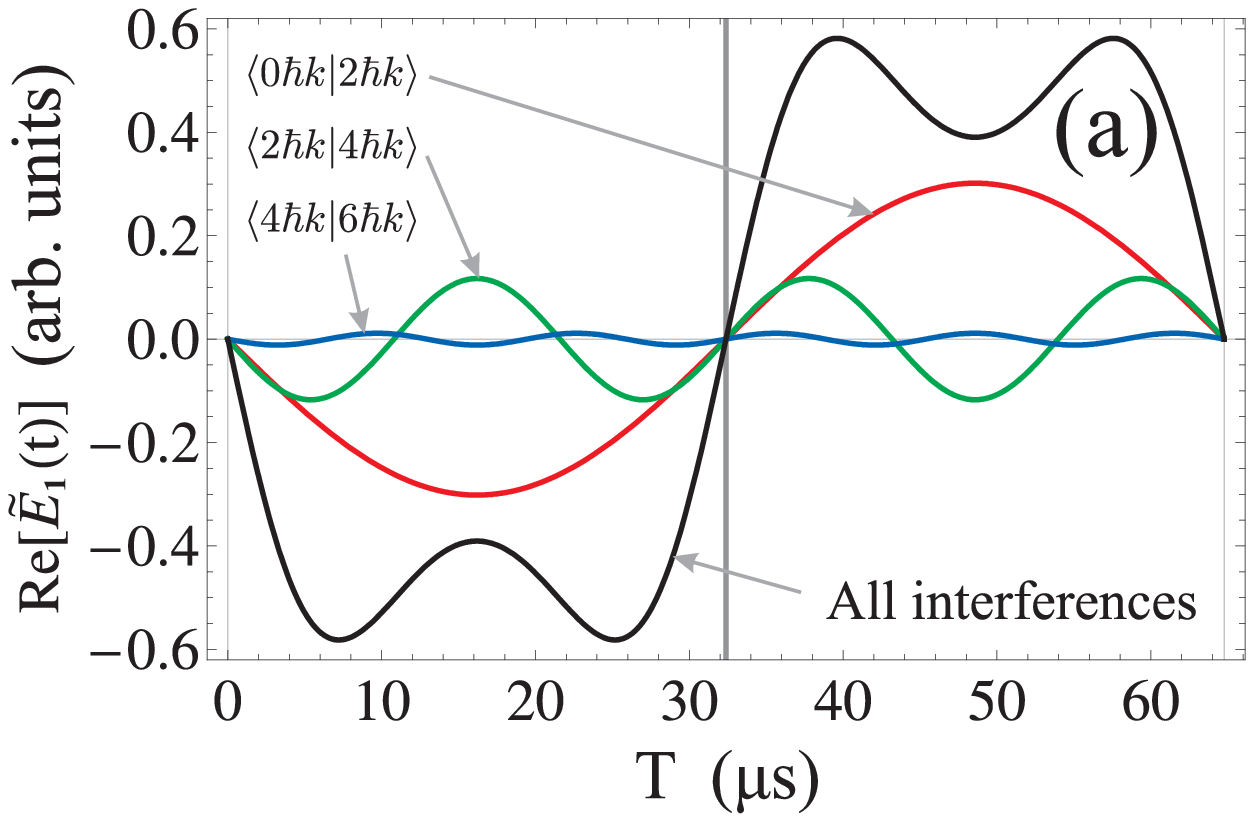}}
  \subfigure{
    \label{fig:ReE1-Components-wSE}
    \includegraphics[width=0.38\textwidth]{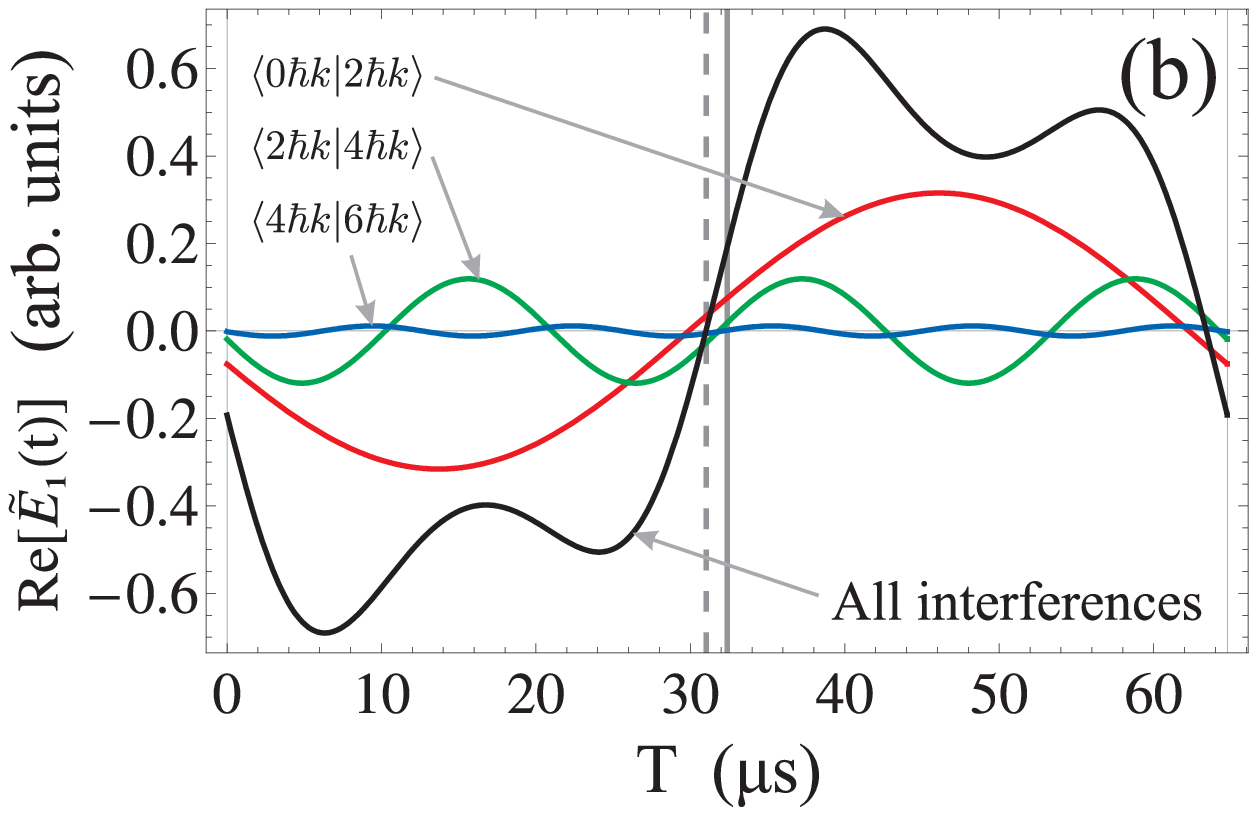}}
  \caption{(Color online) Real part of the harmonics comprising the back-scattered electric field, $\tilde{E}_1(t)$ (\Eq \ref{eqn:E1-Theory-sum}), for a system without spontaneous emission (a: $\theta = 0$) and with spontaneous emission (b: $\theta = -0.133$ rad). Each graph shows harmonics associated with the interference of momentum states $\ket{n\hbar q}$ and $\ket{(n+1)\hbar q}$ for $n = 0, 1$ and 2 (red, green and blue curves, respectively). These harmonics correspond to the $n = 0, 1$ and 2 terms in \Eq \ref{eqn:E1-Theory-sum}. The total back-scattered field, shown as the black curve, is the sum of all harmonics due to the interference between $p$-states differing by $\hbar q$. In (b), the total scattered field is shifted by a temporal amount $\delta t = \theta/\omega_q$ as shown by the dashed grid line. Note that the imaginary part of $\tilde{E}_1(t)$ sums to zero in both (a) and (b) when all terms in \Eq \ref{eqn:E1-Theory-sum} are included. The pulse parameters are the same as in \Fig \ref{fig:SE-Effects}.}
  \label{fig:E1-Components}
\end{figure}

For the moment we consider a system without spontaneous emission ($\theta = 0$ and the complex pulse area, $\Theta_1 \equiv u_1 e^{i\theta}$, is real). The signal develops multiple peaks within each period for large pulse areas (compare \Fig \ref{fig:LowPulseArea-E1} to \Fig \ref{fig:HighPulseArea-E1}). This structure is due to the interference of higher order momentum states which become more populated as $u_1$ increases. We demonstrate this interference by truncating the infinite sum in \Eq \ref{eqn:E1-Theory-sum} to only $n = -2, \ldots, 2$ so that:
\be
\begin{split}
  \label{eqn:E1-approx1}
  \lefteqn{i \tilde{E}_1(t) \approx - J_2(\Theta_1) J_1(\Theta_1^*) e^{-i3\omega_q t}} \\
  & - J_1(\Theta_1) J_0(\Theta_1^*) e^{-i\omega_q t} + J_0(\Theta_1) J_1 (\Theta_1^*) e^{i\omega_q t} \\
  & + J_1(\Theta_1) J_2(\Theta_1^*) e^{i3\omega_q t} + J_2(\Theta_1) J_3(\Theta_1^*) e^{i5\omega_q t}.
\end{split}
\ee
This expression contains the first three harmonics of the back-scattered electric field that oscillate with frequencies $\pm \omega_q$, $\pm 3\omega_q$ and $5\omega_q$, respectively. The sum of different harmonics gives rise to constructive and destructive interference in the field amplitude. The result is a diminished amplitude between the zero crossings of the total back-scattered field---producing the double-peaked structure. This effect is shown in \Fig \ref{fig:ReE1-Components-NoSE} using the terms corresponding to $n = 0$, 1, and 2 in \Eq \ref{eqn:E1-approx1}. In the absence of any phase shifts of the individual harmonics, the shape of the recoil signal ($\tilde{s}_1(t) = |\tilde{E}_1(t)|^2$) is symmetric (black curve in \Fig \ref{fig:SE-Effects}).

We now consider the case where spontaneous emission is present in the system ($\theta \not= 0$). The spontaneous emission parameter, $\theta$, causes a phase shift of each harmonic comprising $\tilde{E}_1(t)$ in \Eq \ref{eqn:E1-Theory}. This is demonstrated by expanding the Bessel functions in \Eq \ref{eqn:E1-approx1} in a Taylor series and keeping only the leading terms:
\bea
  \label{eqn:E1-approx2}
  \tilde{E}_1(t)
  & \approx & -\left(u_1 + \frac{u_1^5}{32}\right) \sin(\omega_q t-\theta) + \frac{u_1^3}{4} \sin(\omega_q t+\theta) \nonumber \\
  & + & \frac{u_1^3}{8} \sin(\omega_q t-3\theta) - \frac{u_1^3}{8} \sin(3\omega_q t-\theta) \\
  & + & \frac{u_1^5}{64} \sin(3\omega_q t+\theta) + \frac{u_1^5}{96} \sin(3\omega_q t-3\theta). \nonumber
\eea
Each term in \Eq \ref{eqn:E1-approx2} has been phase shifted by an integer multiple of $\theta$, but no two terms are shifted by the same temporal amount (colored curves in \Fig \ref{fig:ReE1-Components-wSE}). When all values of $n$ are included, the different temporal shifts sum to a net shift of $\delta t = \theta/\omega_q$ (shown by the dashed grid lines in \Figs \ref{fig:SE-Effects} and \ref{fig:ReE1-Components-wSE}).

Figure \ref{fig:ReE1-Components-wSE} illustrates the harmonic components of the back-scattered field amplitude ($\tilde{E}_1(t)$ from \Eq \ref{eqn:E1-Theory}) associated with the interference $\braket{n\hbar q}{(n+1)\hbar q}$ for $n = 0$, 1 and 2 using $\theta = -0.133$ rad. When summed over all harmonics, the field exhibits more constructive interference on one side of the waveform than the other (black curve in \Fig \ref{fig:ReE1-Components-wSE}) giving rise to an asymmetry in the peak amplitude within each recoil period and a temporal shift of the zeroes. The corresponding field intensity, $\tilde{s}_1(t)$, is shown in \Fig \ref{fig:SE-Effects} as the gray curve. When $\theta = 0$, there is no temporal shift, thus no asymmetry, and we recover the black curve in \Fig \ref{fig:SE-Effects}.

In the experiment, we always observe a negative asymmetry, which means that the amplitude of the first peak of the recoil signal is higher than the second peak between consecutive zeroes, as shown in \Figs \ref{fig:Data-Theory-mLevels} and \ref{fig:Feb24-SimDataOverlay}. A positive asymmetry corresponds to the second peak amplitude being larger than the first. A negative asymmetry implies that the zeroes of the signal are shifted towards earlier times ($\theta, \delta t < 0$), which is a consequence of using a positive detuning. These observations are consistent with theoretical predictions.

The phases of each harmonic in \Eq \ref{eqn:E1-approx2} represent the difference in phase between two interfering momentum states. This phase difference is a direct result of the decay of the excited state into the ground state. When the harmonics are summed over, the relative phases are effectively averaged and the resulting phase shift of the scattered field amplitude represents a weighted average phase difference between all interfering momentum states that differ by $\hbar q$. To first order, this phase is approximately equal to $\theta$. Spontaneous emission is evidently one mechanism by which the relative phases of two interfering momentum states can be non-zero. In the next section, we will show that stimulated emission and absorption produce similar effects.

\subsubsection{Stimulated emission and absorption}
\label{sec:Results-Stim}

\begin{figure}[!t]
  \subfigure{
    \label{fig:PhaseOscillations-1}
    \includegraphics[width=0.232\textwidth]{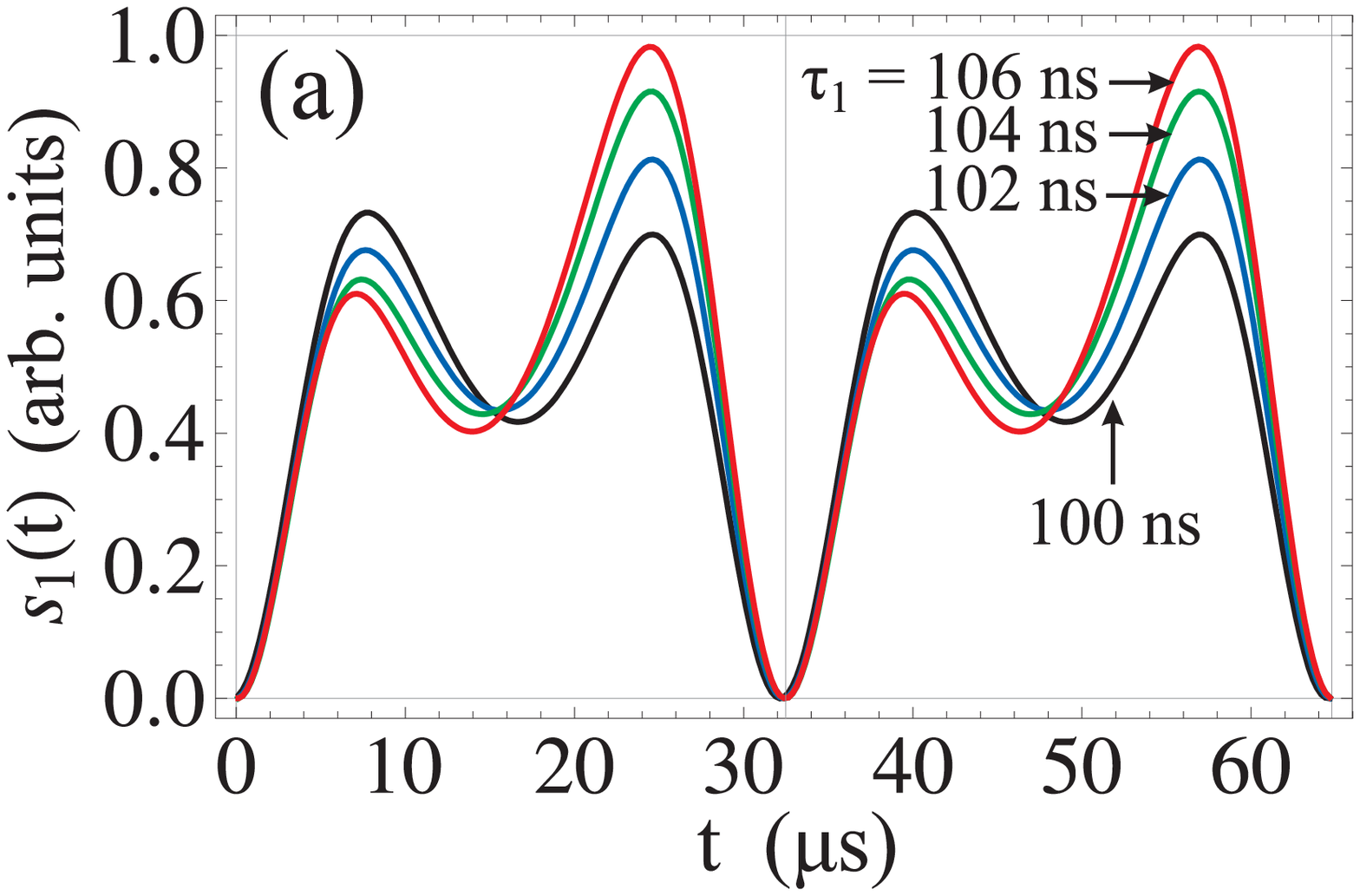}}
  \subfigure{
    \label{fig:gPopulation-1}
    \includegraphics[width=0.232\textwidth]{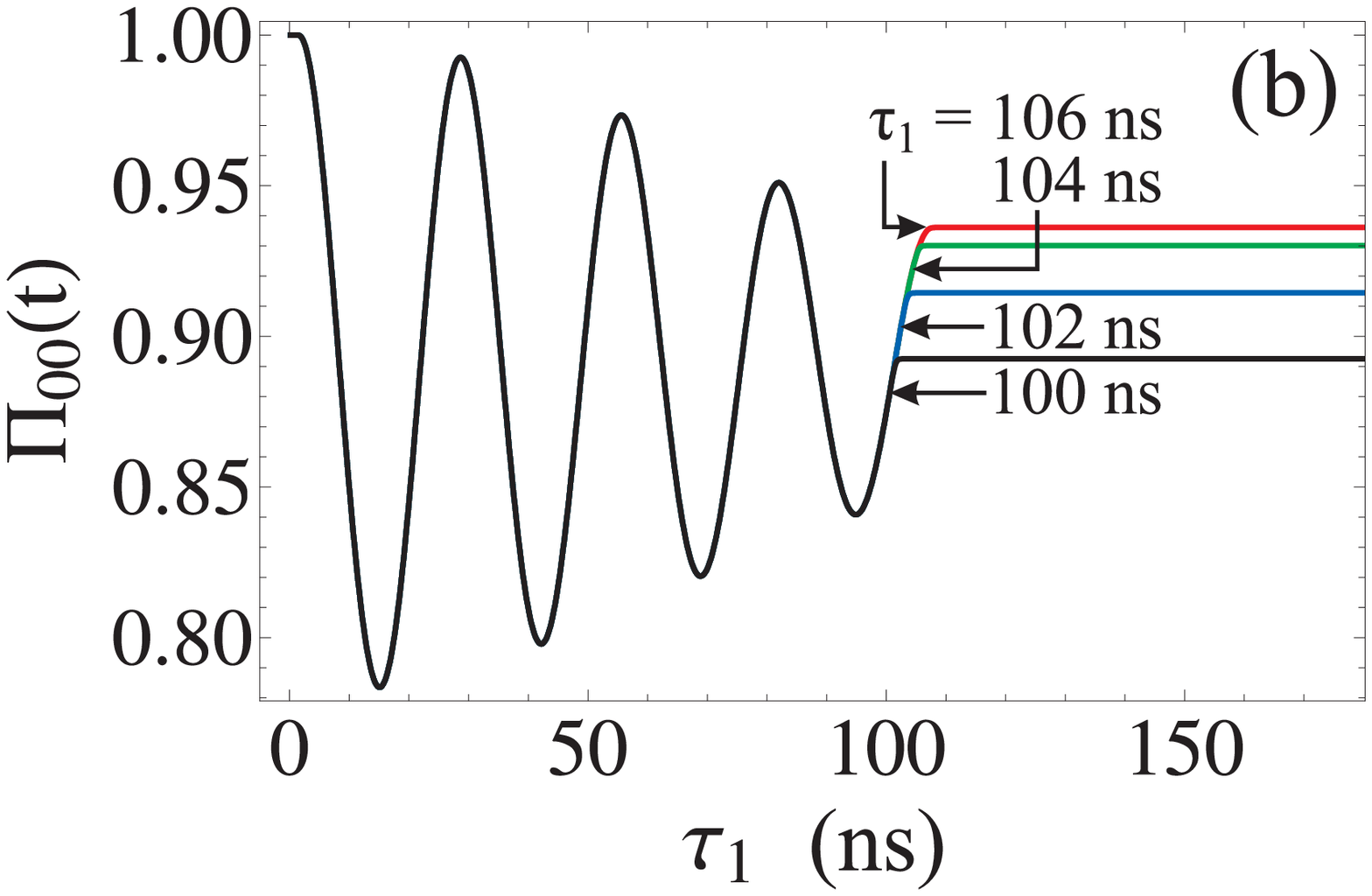}}
  \subfigure{
    \label{fig:PhaseOscillations-2}
    \includegraphics[width=0.232\textwidth]{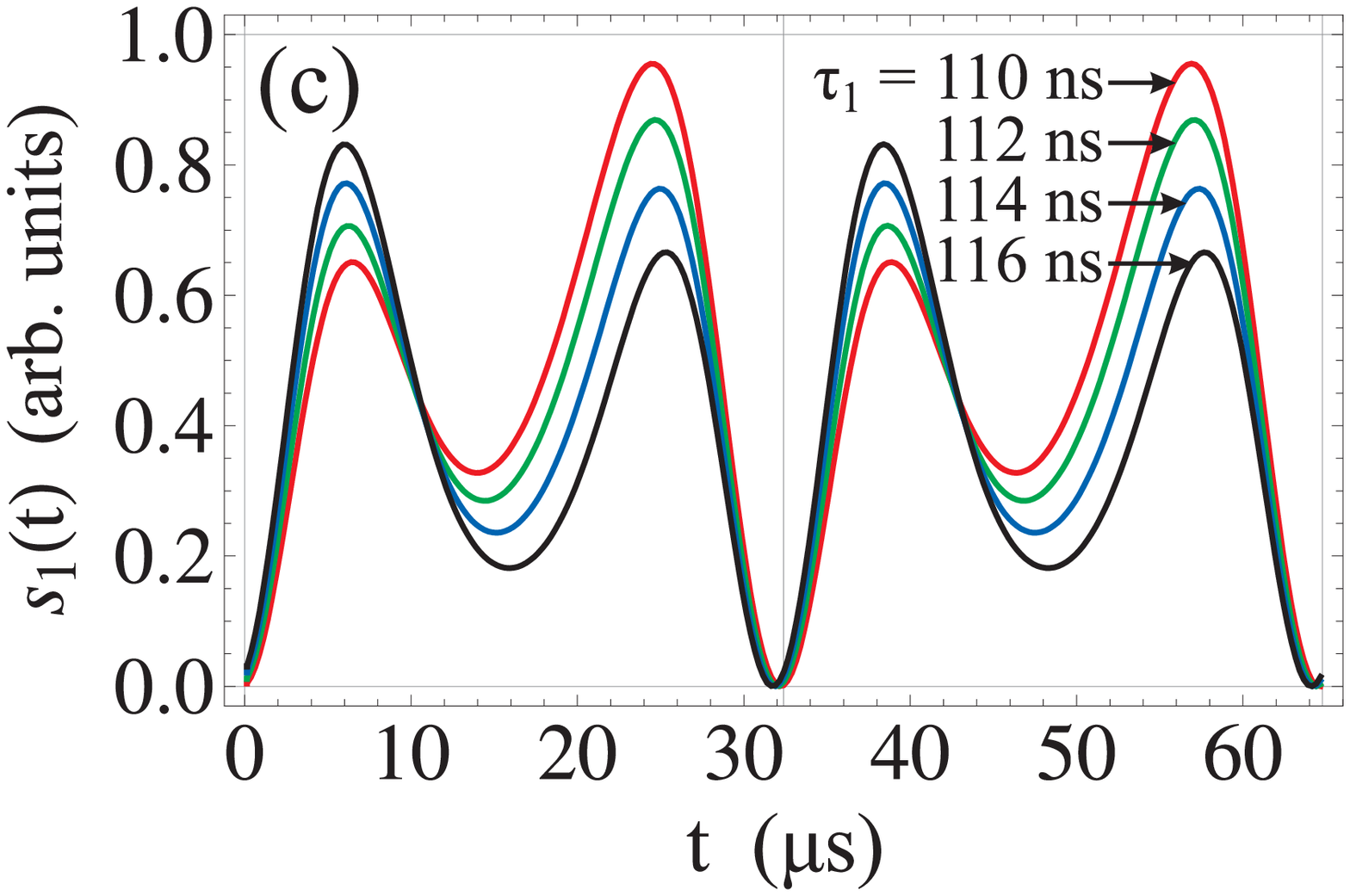}}
  \subfigure{
    \label{fig:gPopulation-2}
    \includegraphics[width=0.232\textwidth]{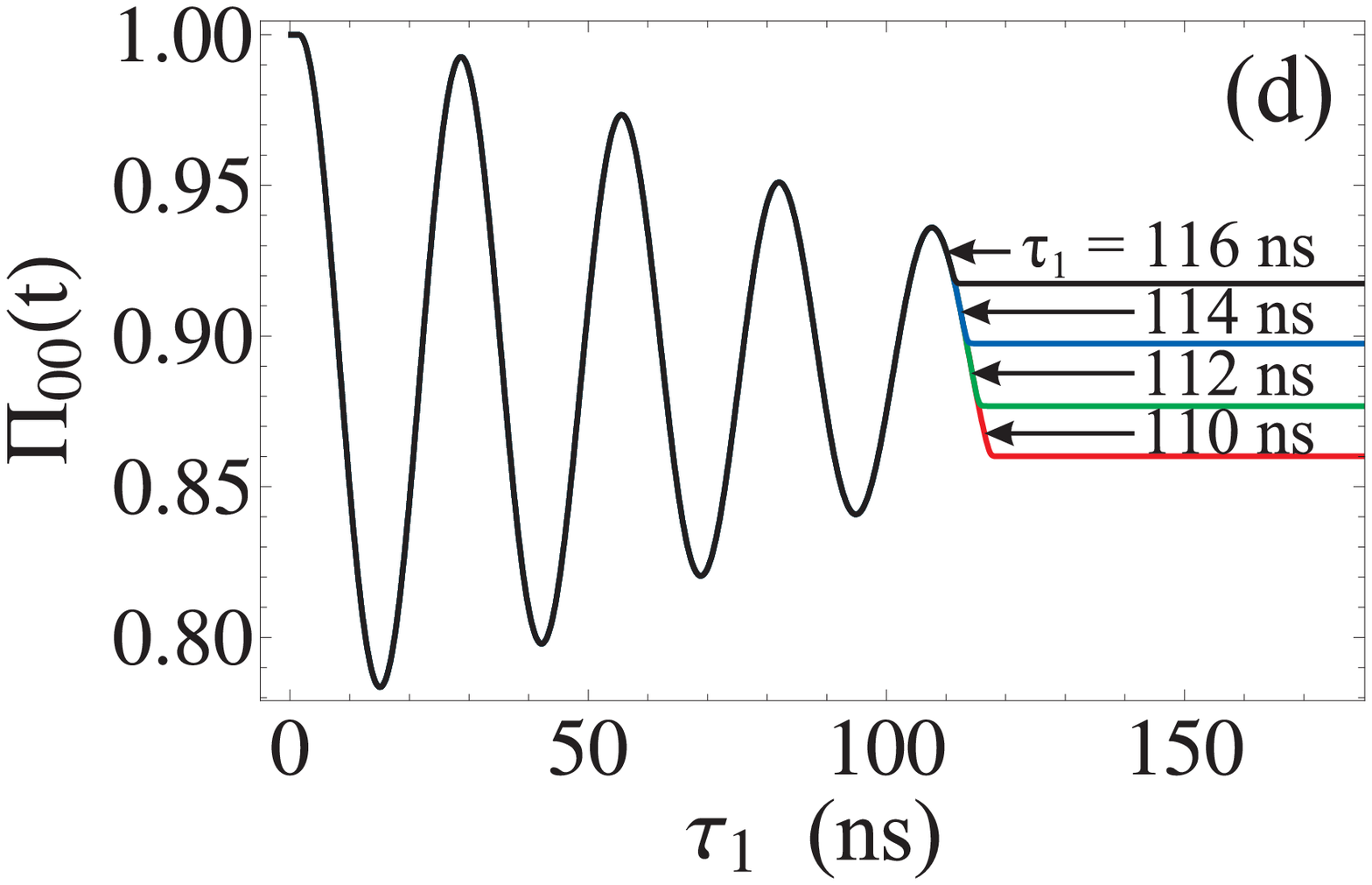}}
  \caption{(Color online) Simulations of one-pulse recoil signal, $s_1(t)$, in the absence of spontaneous emission for pulse durations varying within one Rabi oscillation ($\tau_1 \sim 100 - 116$ ns). In part (a) the recoil signal begins roughly symmetric at $\tau_1 = 100$ ns. As $\tau_1$ is increased the signal exhibits positive asymmetry and shifts toward later times. Part (b) shows the corresponding Rabi oscillation in the ground state population for each $\tau_1$ in (a). In part (c) the signal shifts toward earlier times and shows negative asymmetry as $\tau_1$ increases from 110 to 116 ns. Part (d) shows the corresponding Rabi oscillations for (c). Here, the generalized Rabi frequency is $\sqrt{\Omega_0^2 + \Delta^2} \sim 5.4 \, \Gamma_{\rm{N}} \sim 32$ MHz. Pulse parameters: $\tau_{\rm{rise}} \sim 1$ ns; $\Delta = 5\,\Gamma_{\rm{N}}$; $\Omega_0 = 2\,\Gamma_{\rm{N}}$; $\Gamma = 0$.}
  \label{fig:PhaseOscillations}
\end{figure}

In this section, we illustrate that stimulated emission and absorption between the ground and excited levels leads to a time-dependent asymmetry in the recoil signal. To isolate this effect, we generate simulations of the one-pulse recoil signal in the absence of spontaneous emission (i.e. $\Gamma = 0$). Under these conditions, the signal exhibits either positive or negative asymmetry depending on the pulse duration, $\tau_1$, as shown in \Fig \ref{fig:PhaseOscillations}. This originates from the dynamic evolution of the atomic state amplitudes during the sw pulse (Rabi oscillations) due to stimulated emission and absorption. In contrast, the analytical theory for the one-pulse recoil signal (\Eq \ref{eqn:s1-Theory}), which ignores the dynamic exchange of population between the ground and excited states (see \Eq \ref{eqn:ae}), predicts that there should be no asymmetry or phase shift when there is no spontaneous emission. Results from the simulations indicate that, by changing $\tau_1$, the asymmetry oscillates between positive and negative at the same frequency as the generalized Rabi frequency.

The oscillations in the asymmetry shown in \Fig \ref{fig:PhaseOscillations} originate from the time-dependence of the relative phase between interfering momentum states. As the interaction time with the sw pulse increases, there is a transfer of momentum from the ground state to the excited state and back to the ground state again. The intermediate process of going through the excited state imprints a momentum-dependent phase on the ground state wave function. As a result, the phase difference between two adjacent $p$-states changes as a function of the interaction time, $\tau_1$.

The amplitude of the phase oscillation is largely dependent on the bandwidth of the sw pulse, which has contributions from the pulse duration, $\tau_1$, and the rise time, $\tau_{\rm{rise}}$. A large bandwidth pulse produces a large amplitude phase oscillation. As $\tau_1$ increases, the bandwidth of the pulse decreases---giving rise to a damping of the oscillation amplitude. This result can also be inferred from the time-dependence of the excited state population, since the relative phase between momentum states is a measure of its contribution to the recoil signal.

\subsubsection{Combined phase effects}
\label{sec:Results-DynamicPhase}

In the experiment, both stimulated and spontaneous processes contribute to phase shifts of the recoil signal. In the following section, we discuss the combination of these two mechanisms and measure their affect on the signal. This constitutes one of the most important results in this paper.

We first describe an analytical calculation that predicts a dynamic phase for each momentum state that depends on both the parameter $\theta$ (\Eq \ref{eqn:theta}), which is a measure of the number of spontaneous emission events, and the pulse area $u_1$, which is a measure of the number of stimulated processes. This calculation is valid for the one-pulse recoil signal.

Since the interferometer is sensitive to the relative phase of interfering momentum states, this $p$-dependent phase provides a direct connection between matter wave interference and transitions due to both spontaneous and stimulated processes. However, the calculation of the relative phase is subject to the limitations of the theoretical framework developed in \S\ref{sec:Theory} (weak, far off-resonance pulses and short interaction times). This necessitates the use of simulations to develop more accurate predictions.

When spontaneous emission is included in the theory, the excited state imprints a momentum-dependent phase, $\phi(p)$, on the ground state wave function, which can be obtained from
\be
  \phi(p) = \mbox{arg}\big[ a_g(p) \big],
\ee
where $a_g(p)$ is the amplitude of state $\ket{p}$ in the ground level. The phase of a discrete momentum state $\ket{p = n \hbar q}$ can be obtained using \Eq \ref{eqn:ag(p)} for $a_g(p)$:
\be
  \phi^{(1)}_n = \mbox{arg}\big[ (-i)^n J_n(u_1 e^{i\theta}) \big],
\ee
where the superscript $(1)$ indicates that this calculation is valid for a single sw pulse and the subscript $n$ labels the momentum state. Expanding the Bessel function in a series in powers of $u_1 e^{i\theta}$, one can show that
\begin{align}
\begin{split}
  \label{eqn:phi1_n}
  & \tan \left( \phi^{(1)}_n + n\frac{\pi}{2} \right) = \\
  & \;\;\;\;\;
  \frac{\ds \sum_{j = 0}^{\infty} \frac{(-1)^j}{j!(j+|n|)!}\left(\frac{u_1}{2}\right)^{2j+|n|} \sin[(2j+|n|)\theta]}
       {\ds \sum_{l = 0}^{\infty} \frac{(-1)^l}{l!(l+|n|)!}\left(\frac{u_1}{2}\right)^{2l+|n|} \cos[(2l+|n|)\theta]}.
\end{split}
\end{align}
To first order: $\phi^{(1)}_n \approx |n|\theta - n \pi/2$, which is independent of the pulse area, $u_1$, and therefore ignores stimulated processes. This result indicates that the phase difference between interfering momentum states is
\be
  \label{eqn:DeltaPhi1n}
  \Delta \phi^{(1)}_n = \phi^{(1)}_n - \phi^{(1)}_{n-1} \approx \left\{
  \begin{array}{cr} \;\;\, \theta + \pi/2 & \mbox{for } n \ge 1, \\ -\theta + \pi/2 & \mbox{for } n < 1. \end{array}
  \right.
\ee
For higher orders, however, stimulated processes contribute to the phase of each momentum state and there is a distinct dependence on the pulse area. The AI is sensitive to the average relative phase between $p$-states that differ by $\hbar q$. For an arbitrary ground state amplitude $a_g(p)$, this quantity can be expressed as
\be
  \label{eqn:DeltaPhi}
  \langle \Delta \phi \rangle = \mbox{arg}\left[ \int a_g(p) a_g^*(p - \hbar q) dp \right],
\ee
where the $\langle \cdots \rangle$ notation denotes an average over momentum. For the specific case of the one-pulse recoil experiment, the average relative phase is
\be
  \label{eqn:DeltaPhi1}
  \langle \Delta \phi^{(1)} \rangle
  = \mbox{arg}\left[ - i \sum_{n} J_n \big( u_1 e^{i\theta} \big) J_{n-1} \big( u_1 e^{-i\theta} \big) \right],
\ee
which has a distinct dependence on the sw pulse duration, $\tau_1$.

\begin{figure}[!t]
  \includegraphics[width=0.4\textwidth]{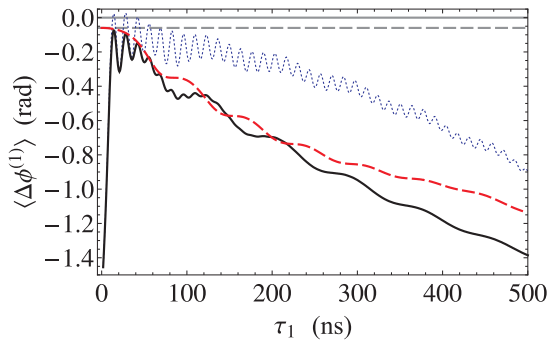}
  \caption{(Color online) Average relative phase between momentum states differing by $\hbar q$ as a function of sw pulse duration, $\tau_1$. An offset of $\pi/2$ has been added to the vertical axis. The dashed gray grid line shows the value of $\theta \sim -0.06$ rad from \Eq \ref{eqn:theta}. The dashed red line is a calculation based on \Eq \ref{eqn:DeltaPhi1} with $u_1$ given by \Eq \ref{eqn:u1}. The dotted blue line shows the prediction of \Eq \ref{eqn:DeltaPhi} computed from a simulation in the absence of spontaneous emission. The solid black line is the analogous prediction of a simulation including spontaneous emission. The steep rise of the two simulations at small $\tau_1$ represents the first Rabi cycle of phase transfer from $p = 0$ to neighboring states. These states initially have zero amplitude and their phases are not well defined. Pulse parameters: $\tau_{\rm{rise}} \sim 1$ ns; $\Delta = 8.4\,\Gamma_{\rm{N}}$; $\Omega_0 = 4.5\,\Gamma_{\rm{N}}$.}
  \label{fig:AvgRelativePhase}
\end{figure}

Figure \ref{fig:AvgRelativePhase} illustrates the dependence of $\langle \Delta \phi^{(1)} \rangle$ on the sw pulse duration, $\tau_1$. Predictions are shown for both theory (dashed red line---\Eq \ref{eqn:DeltaPhi1}) and simulations in the presence and absence of spontaneous emission (dotted blue and solid black lines, respectively---\Eq \ref{eqn:DeltaPhi}).

For small $\tau_1$, the theoretical expression for the average relative phase (\Eq \ref{eqn:DeltaPhi1}) is approximately equal to $\theta$. This result can also be inferred from \Eq \ref{eqn:DeltaPhi1n}, which is valid only for small $u_1$. According to theory, the contribution from spontaneous emission is a constant phase offset equal to $\theta$ (\Eq \ref{eqn:theta}). As the interaction time increases, the theory predicts an increase in $|\langle \Delta \phi^{(1)} \rangle|$, as well as a slow oscillation. The frequency of this oscillation is empirically determined to be $\sim 2 \Omega_0^2/\Delta \sim 2.7$ MHz, which is due to stimulated emission and absorption.

Results of simulations in the absence of spontaneous emission (dotted blue line) show an additional high frequency oscillation in $\langle \Delta \phi^{(1)} \rangle$ as a function of $\tau_1$. These oscillations are caused by the transfer of phase between the ground and excited state due to stimulated processes, and have a frequency approximately equal to the generalized Rabi frequency $\sqrt{\Omega_0^2 + \Delta^2} \sim 56$ MHz. These high frequency phase oscillations are not taken into account in \Eq \ref{eqn:DeltaPhi1} as a result of the far off-resonance assumption. When spontaneous emission is included in the simulations (solid black line), the oscillations are highly damped. As the pulse area increases, there is a significant departure of the theory from the simulations. We attribute this departure to the absence of dynamic population transfer between the ground and excited states and the weak-field assumption made in the theory.

\begin{figure}[!t]
  \includegraphics[width=0.4\textwidth]{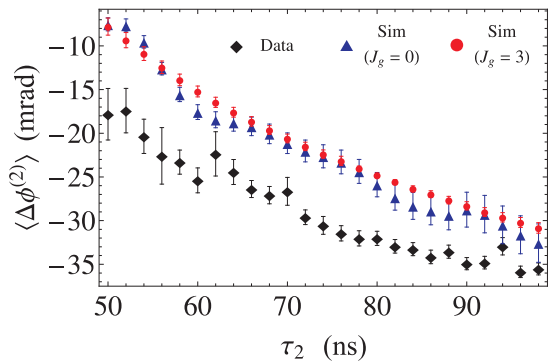}
  \caption{(Color online) Variation of average relative phase, $\langle \Delta\phi^{(2)} \rangle$, as a function of $2^{\rm{nd}}$ pulse length. The relative phase measured from fits to experimental data ($\blacklozenge$) are overlayed with the results of a two-level simulation ($\blacktriangle$, where $J_g = 0$) and a multi-level simulation ($\bullet$, where $J_g = 3$). A subset of the multi-level simulations overlayed with data is shown in \Fig \ref{fig:Feb24-SimDataOverlay}. $\langle \Delta\phi^{(2)} \rangle$ is measured from the data and the multi-level simulation by fitting to the square of \Eq \ref{eqn:E2-mg-sum} and using the best fit value of $\theta$. Similarly, it is measured for the two-level simulation by fitting to \Eq \ref{eqn:s2-Theory}. A subset of the data from which $\langle \Delta\phi^{(2)} \rangle$ is extracted is shown in \Fig \ref{fig:Data-Theory-mLevels}, with the corresponding fits shown as solid red curves. The error bars for each set of points are equal to the statistical uncertainties in $\theta$ from the corresponding fits. The pulse parameters are the same as those in \Fig \ref{fig:Data-Theory-mLevels}, except for the two-level simulations where the rise time was reduced to $\tau_{\rm{rise}} = 10$ ns to illustrate the presence of oscillations in $\langle \Delta\phi^{(2)} \rangle$ as a function of $\tau_2$.}
  \label{fig:Data-Sims-theta}
\end{figure}

We now review results of measurements of the average relative phase, $\langle \Delta \phi^{(2)} \rangle$, from the two-pulse recoil experiment and corresponding simulations. Figure \ref{fig:Data-Sims-theta} shows the variation in $\langle \Delta \phi^{(2)} \rangle$ as a function of $\tau_2$. This variation corresponds to the change in the asymmetry of the signal shapes shown in \Fig \ref{fig:Data-Theory-mLevels}. We measure $\langle \Delta \phi^{(2)} \rangle$ from the data by fitting to the square of \Eq \ref{eqn:E2-mg-sum} (which includes effects due to $m$-levels) and using the best fit value of $\theta$. From the data, $\langle \Delta \phi^{(2)} \rangle$ seems to decrease and level out near $-35$ mrad after $\tau_2 \sim 90$ ns. This can be explained by the excited state population approaching steady state---giving rise to a saturation of momentum state excitation.

Figure \ref{fig:Data-Sims-theta} also shows results of two-level simulations ($J_g = 0$) implemented with the same pulse parameters as the data (except for the rise time which was set to $\tau_{\rm{rise}} = 10$ ns). Here, $\langle \Delta \phi^{(2)} \rangle$ is measured by fitting each recoil curve generated by the simulation to \Eq \ref{eqn:s2-Theory} (which incorporates only two levels). This is equivalent to the method used to measure the average relative phase from data. The trend of the two-level simulation is qualitatively similar to the black curve shown in \Fig \ref{fig:AvgRelativePhase}. In comparison to the data we find an additional superimposed oscillation and a constant phase offset. The frequency of this phase oscillation is the same as the generalized Rabi frequency ($\sqrt{\Omega_0^2 + \Delta^2} \sim 59$ MHz in this case). This is consistent with effects due to stimulated emission and absorption, as discussed in \S\ref{sec:Results-Stim}. The oscillations are quickly damped as the excited state population approaches steady state. The rate at which this occurs is mainly determined by the bandwidth of the pulse (which has contributions from both $\tau_2$ and $\tau_{\rm{rise}}$) and the spontaneous emission rate. In these two-level simulations, we have reduced the rise time of the sw pulses from 20 ns (as measured in the experiment) to 10 ns in order to demonstrate the presence of these oscillations.

The red circles shown in \Fig \ref{fig:Data-Sims-theta} represent simulations including magnetic sub-levels ($J_g = 3$). These results show no phase oscillations as a function of interaction time, which is consistent with the data. The absence of phase oscillations in the data and the multi-level simulation is attributed to the relatively long rise time ($\tau_{\rm{rise}} \sim 20$ ns) of the excitation pulses, as discussed in the previous sub-section. There is also additional damping of phase oscillations due to the interference between back-scattered fields from each ground state magnetic sub-level.

Although the trend in $\langle \Delta \phi^{(2)} \rangle$ from both the two-level and the multi-level simulations is similar to the data, there is a small phase offset between the two. This can be explained in the following manner. Firstly, the offset is equivalent to a difference in the level of asymmetry between the data and the simulations. This is not surprising since our estimates of various input parameters, such as the Rabi frequency and $m$-level populations, are not directly measured in the experiment. Secondly, our ability to measure $\langle \Delta \phi^{(2)} \rangle$ from the data is limited by the signal-to-noise. Since $\langle \Delta \phi^{(2)} \rangle$ is estimated by fitting to the data, and each point in the fit is equally weighted, this results in an over-estimation of the phase value and an under-estimation in the uncertainty of that value. This is especially true at small $\tau_2$. This is because the quantity being measured is highly shape-dependent and, at small $\tau_2$, where there is very little asymmetry in the data, a small amount of noise can cause a large error in the measurement.

The time evolution of the relative phase predicted by the two-level and the multi-level simulations, however, are very similar. This suggests that $\langle \Delta \phi^{(2)} \rangle$ is largely unaffected by the presence of magnetic sub-levels, but depends mostly on the sw pulse parameters and the magnitude of the spontaneous emission rate---which, aside from the pulse rise time, were the same in both simulations. This is consistent with the theory presented above.

In a previous paper \cite{Beattie3}, we investigated the influence of spontaneous emission on the two-pulse recoil signal using both theory and experiments. We used a phenomenological model for the two-pulse recoil signal that was successful in predicting the asymmetry in the signal shape and the temporal offset. Since the spontaneous emission parameter, $\theta$, is related to the spontaneous emission rate, $\Gamma$, it appeared that there was a variation in $\Gamma$ with the pulse duration and a measurement of $\Gamma$ appeared to be possible by fitting experimental data using this model (\Eq 26 in \Ref \cite{Beattie3}). However, the average relative phase of the signal, $\langle \Delta \phi^{(2)} \rangle$, depends on both $\theta$ and the pulse area. The phenomenological model did not take into account this pulse area dependence. As a result, erroneous values of $\theta$, and therefore $\Gamma$, were measured when $\tau_2$ changed. Thus, the conclusion that $\Gamma$ depended on $\tau_2$ was not justified. The dependence of the average relative phase on spontaneous and stimulated processes, as well as effects due to magnetic sub-levels, was not fully understood until recently. With this interpretation, we can explain the variation of $\theta$ with $\tau_2$ observed in \Ref \cite{Beattie3}.

\section{Conclusion}
\label{sec:Conclusion}

We have developed a comprehensive numerical simulation of a time-domain echo interferometer that is successful in predicting a variety of physical effects associated with matter wave interference. The simulation accurately models experimental results for measured input parameters. Based on these results, we have determined that the interference between back-scattered electric fields from each ground state magnetic sub-level significantly affects the signal shape, which is consistent with experimental observations. Additionally, a more complete understanding of the phase modulation of the atomic wave function induced by standing wave pulses has been developed on the basis of the simulation.

The simulation motivated the development of an improved analytical model of the signal that accounts for both spontaneous emission and ground state magnetic sub-levels. Using the analytical model, we were able to improve the relative error in measurements of the recoil frequency by a factor of $\sim 3$. We were also able to estimate magnetic sub-level populations at the time of the read-out pulse.

Through theoretical analysis and simulations, we have clarified the role of spontaneous emission on the recoil signal. We have also explained how the relative phase between interfering momentum states affects the signal on the basis of stimulated emission and absorption, which is most important when the excited state is significantly populated. These effects are not unique to the AI considered in this paper, but apply more generally to interferometers that involve momentum state interference induced by laser excitation, such as frequency domain AIs \cite{Weel}.

The robustness of the simulations suggest that they can be used to study the AI under a variety of interesting conditions, such as the near-resonance ($|\Delta| \sim \Gamma$), high intensity ($\Omega_0 \sim |\Delta|$) or long-pulse ($\tau_j \gg |\Delta|^{-1}$) regimes. An important application of this work also includes the study of systematic effects on a precision measurement of the atomic recoil frequency using the three-pulse technique outlined in \Refs \cite{Beattie2,Beattie3}.

\begin{acknowledgments}

This work was supported by the Canada Foundation for Innovation, Ontario Innovation Trust, Natural Sciences and Engineering Research Council of Canada, Ontario Centres of Excellence and York University. I. Yavin is supported by the James Arthur fellowship. This work was made possible by the facilities of the Shared Hierarchical Academic Research Computing Network (SHARCNET:www.sharcnet.ca) and Compute/Calcul Canada. We would also like to thank Marko Horbatsch and Randy Lewis of York University and Paul Berman of the University of Michigan for helpful discussions.

\end{acknowledgments}
\appendix*
\renewcommand{\theequation}{A\arabic{equation}} 
\setcounter{equation}{0}  

\section*{Appendix}

We review the calculation of the recoil signal in the one-pulse regime. The evolution of the atomic state amplitudes is governed by the Schr\"{o}dinger equation
\be
  \tilde{\mathcal{H}} \left( \begin{array}{c} a_e \\ a_g \end{array} \right)
  = i \hbar \left( \begin{array}{c} \dot{a}_e \\ \dot{a}_g \end{array} \right),
\ee
where $\tilde{\mathcal{H}}$ is the Hamiltonian given by \Eq \ref{eqn:H-1}. We assume that the rate of change of the excited state amplitude is small compared to the detuning ($|\dot{a}_e| \ll |\Delta|$), which results in the following relations:
\begin{subequations}
\bea
  \label{eqn:ae}
  a_e & = & \frac{\Omega e^{i\theta}}{(\Delta^2 + \gamma^2)^{1/2}} \, a_g, \\
  \label{eqn:ag-dot}
  \dot{a}_g & = & -i \frac{\Omega^2 e^{i\theta}}{(\Delta^2 + \gamma^2)^{1/2}} \, a_g.
\eea
\end{subequations}
Here, $\theta$ is a phase that characterizes the amount of spontaneous emission occurring during the pulse and is given by \Eq \ref{eqn:theta}. Equation \ref{eqn:ag-dot} is integrated to obtain the ground state amplitude for a pulse of duration $\tau_1$
\be
  \label{eqn:ag(r)}
  a_g(\bm{r}) = a_g(\bm{r},0) e^{-i \Theta_1 \cos \bm{q}\cdot\bm{r}},
\ee
where $\bm{q} = \bm{k}_1 - \bm{k}_2 = 2 \bm{k}$ is the difference between the $k$-vectors comprising the sw and $\Theta_1 \equiv u_1 e^{i\theta}$ is the complex pulse area with magnitude $u_1$ given by
\be
  \label{eqn:u1}
  u_1 \equiv \frac{\Omega_0^2 \tau_1}{2 \Delta} \left[1 + \left(\frac{\Gamma}{2\Delta}\right)^2 \right]^{-1/2}.
\ee
We assume the initial state of the atom to be a plane wave, $a_g(\bm{r},0) = V^{-1/2} e^{i \bm{p}_0 \cdot \bm{r}/\hbar}$, with momentum $\bm{p}_0 = M \bm{v}_0 = \hbar \bm{k}_0$ and interaction volume $V$.

In \Eq \ref{eqn:ag(r)} we have ignored a $-i\Theta_1$ term in the exponent because it is independent of $\bm{r}$ and is therefore unimportant for interference. It also causes $a_g$ to decay exponentially during the pulse (since $\Theta_1$ is a complex quantity), which does not preserve normalization. Since we assume the system is closed (no atoms are being lost) this corresponds to an unwanted physical process and is a direct result of using a Hamiltonian that is non-Hermitian (\Eq \ref{eqn:H-1}).

From \Eq \ref{eqn:ag(r)}, it is clear that the effect of interacting with the sw pulse is to spatially modulate the phase of the ground state amplitude. Equivalently, the standing wave field is a phase grating that diffracts the atomic wave function into a superposition of momentum states. Using the Jacobi-Anger expansion, we decompose the ground state amplitude into its harmonics:
\be
  \label{eqn:ag(r)-2}
  a_g(\bm{r}) = a_g(\bm{r},0) \sum_{n=-\infty}^{\infty} (-i)^n J_n(\Theta_1) e^{in\bm{q}\cdot\bm{r}},
\ee
where $J_n(x)$ is the $n^{\rm{th}}$ order Bessel function of the $1^{\rm{st}}$ kind. Here, the index $n$ indicates the order of each harmonic, which corresponds to a specific momentum state: $e^{in\bm{q}\cdot\bm{r}} \propto \ket{\bm{p} = n\hbar\bm{q}}$. The sw pulse only populates the ground state with even multiples of the photon momentum, $\hbar \bm{k}$. In contrast, the excited state is populated with only odd multiples of $\hbar \bm{k}$ due to an extra factor of $\Omega(\bm{r})$ in the excited state amplitude, $a_e$, which can be written as:
\begin{align}
\begin{split}
  \label{eqn:a_e(r)}
  & a_e(\bm{r}) = \frac{\Omega_0 e^{i\theta}}{2(\Delta^2 + \Gamma^2)^{1/2}} \, a_g(\bm{r},0) \\
  & \;\;
  \times \sum_n (-i)^n J_n(\Theta_1) \left[ e^{i(2n-1)\bm{k}\cdot\bm{r}} + e^{i(2n+1)\bm{k}\cdot\bm{r}} \right].
\end{split}
\end{align}
The population of each momentum state, $|J_n(\Theta_1)|^2$, is completely determined by the magnitude of the pulse area, $u_1$. However, the phase of each momentum state is determined by both the magnitude \emph{and} the phase of the pulse area, $u_1$ and $\theta$, respectively. This has an important consequence for the interference of two momentum states (see \S\ref{sec:Results-DynamicPhase}).

After the sw pulse turns off, the atom evolves in free space. The population of the excited state is scaled by a factor $\Omega_0/|\Delta| \ll 1$ compared to the ground state, and is therefore ignored. The Hamiltonian in free space contains only the kinetic energy term, $\hat{P}^2/2M$, where $\hat{P}$ is the momentum operator. Therefore, the calculation of the ground state amplitude at time $t$ is much more transparent by transforming into momentum space:
\be
  \label{eqn:ag(p)}
  a_g(\bm{p}) = \frac{(2\pi\hbar)^{3/2}}{\sqrt{V}} \sum_n (-i)^n J_n(\Theta_1) \delta^3(\bm{p} - \bm{p}_0 - n \hbar \bm{q}).
\ee
The solution to the Schr\"{o}dinger equation in this regime dictates that the phase of the ground state amplitude is modulated differently for each momentum state:
\be
  a_g(\bm{p},t) = a_g(\bm{p}) \exp \left[-\frac{i}{\hbar} \frac{p^2}{2M} \, t \right].
\ee
By transforming back to position space, the ground state amplitude a time $t$ after the sw pulse is:
\begin{align}
\begin{split}
  \label{eqn:ag(r,t)}
  & a_g(\bm{r},t) = \frac{1}{\sqrt{V}} \, e^{i(\bm{p}_0 \cdot \bm{r} - \epsilon_0 t)/\hbar} \\
  & \;\;
  \times \sum_n (-i)^n J_n(\Theta_1) e^{i n \bm{q}\cdot\bm{r}} e^{-i n \bm{q}\cdot\bm{v}_0 t} e^{-i n^2 \omega_q t},
\end{split}
\end{align}
where $\epsilon_0 = p_0^2/2M$ is the initial kinetic energy of the atom, $\bm{v}_0 = \bm{p}_0/M$ is the initial velocity and $\omega_q = \hbar q^2/2M$ is the two-photon recoil frequency.

We are interested in how the ground state atomic density distribution, $\rho_g(\bm{r},t) = |a_g(\bm{r},t)|^2$, evolves in time:
\begin{align}
\begin{split}
  \label{eqn:rho(r,t)}
  \rho_g(\bm{r},t) & = \frac{1}{V} \sum_{n,\eta} i^{\eta} J_n(\Theta_1) J_{n+\eta}(\Theta_1^*) \\
  & \times e^{-i\eta \bm{q}\cdot\bm{r}} e^{i\eta \bm{q}\cdot\bm{v}_0 t} e^{i\eta(2n+\eta)\omega_q t}.
\end{split}
\end{align}
This quantity contains a sum over all orders of interference between momentum states and is shown in \Figs \ref{fig:LowPulseArea-DensityPlot} and \ref{fig:HighPulseArea-DensityPlot}. The index $\eta$---equal to the difference between momentum states in multiples of $\hbar \bm{q}$---denotes the order of the interference. This form of $\rho_g$ helps identify the phase due to spontaneous emission and its subsequent role in the one-pulse recoil signal. The density is modulated in time at integer multiples of $\omega_q$. This multiple depends on the order of the momentum state, $n$, and the order of the interference, $\eta$. The Doppler phase $\eta \bm{q}\cdot\bm{v}_0 t$ results in a dephasing of the density modulation when the density is averaged over the velocity distribution of the atomic sample. However, for the purposes of this discussion, we have assumed that the velocity distribution is infinitely narrow such that Doppler dephasing is negligible. As a result, we have ignored the Doppler phase in all expressions that follow.

The density modulation can be detected in an experiment by measuring coherent Bragg scattering from a traveling wave read-out pulse. However, this technique is sensitive only to the $q$-Fourier harmonic of the density distribution when the read-out pulse has the same wavelength, $\lambda$, as the sw pulse. This is because Bragg scattering occurs for structures with a spatial periodicity of $m \lambda/2$, for $m = 1, 2, \dots$. Higher-order spatial harmonics of the density grating (i.e those periodic at $\lambda/4$, $\lambda/6, \ldots$) do not scatter the light. Thus, the back-scattered electric field amplitude is proportional to the $q$-Fourier harmonic of $\rho_g(\bm{r},t)$, which corresponds to the amplitude of the $e^{-i\eta \bm{q} \cdot \bm{r}}$ term in \Eq \ref{eqn:rho(r,t)} with $\eta$ set to unity:
\be
  \label{eqn:E1-Theory-sum}
  \tilde{E}_1(t) \propto \sum_{n=-\infty}^{\infty} i J_n(\Theta_1) J_{n+1}(\Theta_1^*) e^{i(2n+1)\omega_q t}.
\ee

A simpler form of the back-scattered electric field amplitude is realized by using the Bessel function summation theorem \cite{Gradshteyn, Gosset, Beattie3}
\be
\begin{split}
  \sum_n J_n(\Theta_1) J_{n+\eta}(\Theta_1^*) e^{i(2n+\eta)\omega_q t} = \\
  i^{\eta} J_{\eta} (\kappa_1) \left( \frac{\sin(\omega_q t - \theta)}{\sin(\omega_q t + \theta)} \right)^{\eta/2}
\end{split}
\ee
where
\be
  \label{eqn:kappa1}
  \kappa_1 \equiv 2 u_1 \sqrt{\sin(\omega_q t + \theta) \sin(\omega_q t - \theta)},
\ee
resulting in the following expression for $\tilde{E}_1(t)$
\be
  \tilde{E}_1(t) \propto - J_1( \kappa_1 ) \left( \frac{\sin(\omega_q t - \theta)}{\sin(\omega_q t + \theta)} \right)^{1/2}.
\ee
However, this form contains a singularity at $t = -(\theta + m \pi)/\omega_q$, which is not physical. The singularity can be removed by using the recurrence relation: $2 \nu J_{\nu}(x) = x \big[ J_{\nu-1}(x) + J_{\nu+1}(x) \big]$, such that
\be
  \label{eqn:E1-Theory-A}
  \tilde{E}_1(t) \propto -u_1 \sin(\omega_q t - \theta) \big[ J_0( \kappa_1 ) + J_2( \kappa_1 ) \big].
\ee
The analytical expression for the one-pulse recoil signal, denoted by $\tilde{s}_1$, is the intensity of the back-scattered field, which is proportional to $|\tilde{E}_1(t)|^2$:
\begin{align}
\begin{split}
  \label{eqn:s1-Theory}
  \tilde{s}_1(t)
  & \propto u_1^2 \sin^2(\omega_q t - \theta) \big[ J_0( \kappa_1 ) + J_2( \kappa_1 ) \big]^2.
\end{split}
\end{align}
This expression for the recoil signal is valid for an atomic sample with a narrow velocity distribution (BEC conditions) after interacting with a single sw pulse.

\bibliography{WPEPS-Bibliography}

\end{document}